\documentclass[10pt]{elsarticle}
 
\usepackage{calrsfs,graphicx,array,natbib,enumitem}
\usepackage{setspace,wrapfig,marvosym,bbm,stmaryrd,centernot,color,url,amssymb}
\usepackage{amsmath}

\newtheorem{rem}{Remark}

\newlength{\kaka}

\newcommand{\ahref}[2]{}

\newcommand{\obs}{^{\text{obs}}}

\newcommand{\R}{\mathbb{R}}

\newcommand{\ff}{^{\text{\tiny f}}}

\newcommand{\beq}{\begin{equation}}
\newcommand{\eeq}{\end{equation}}
\newcommand{\lb}{\label}

\newcommand{\bea}{\begin{eqnarray}}
\newcommand{\eea}{\end{eqnarray}}
\newcommand{\bxr}{\begin{array}}
\newcommand{\exr}{\end{array}}

\newcommand\exs{\hspace*{0.4mm}}

\newcommand\nxs{\hspace*{-0.2mm}}

\newcommand{\norms}[1]{\parallel\! #1 \!\parallel}

\newcommand{\bC} {\boldsymbol{C}}
\newcommand{\bK} {\boldsymbol{K}}
\newcommand{\bV} {\boldsymbol{V}}
\newcommand{\bR} {\boldsymbol{\sf R}}

\newcommand{\bn} {\boldsymbol{n}}
\newcommand{\ba} {\boldsymbol{a}}

\newcommand{\bx} {\boldsymbol{x}}
\newcommand{\by} {\boldsymbol{y}}

\newcommand{\bg} {{\boldsymbol{g}}}

\newcommand{\bI} {\boldsymbol{I}}

\newcommand{\sip} {\!\cdot\!}

\newcommand{\bzero}{\boldsymbol{0}}

\newcommand{\bu} {\boldsymbol{u}}

\newcommand{\bv} {\boldsymbol{v}}

\newcommand{\bxi} {\boldsymbol{\xi}}

\newcommand{\bw} {\boldsymbol{w}}

\newcommand{\bPhi}{\boldsymbol{\Phi}}

\begin{document}

\begin{frontmatter}

\title{Experimental validation of differential evolution indicators \\ for ultrasonic imaging in unknown backgrounds}

\author{Fatemeh Pourahmadian\corref{cor1}}

\address{Department of Civil, Environmental \& Architectural Engineering, University of Colorado Boulder, USA \\ Department of Applied Mathematics, University of Colorado Boulder, USA}

\cortext[cor1]{Corresponding author: tel. 303-492-2027, email {\tt fatemeh.pourahmadian@colorado.edu}}

\date{\today}

\begin{abstract}

The {differential evolution indicators}, recently introduced~\cite{pour2019} for imaging mechanical evolution in highly scattering solids, is examined in a laboratory setting with the focus on spatiotemporal tracking of an advancing damage zone in an elastic specimen. To this end, a prismatic slab of charcoal granite is quasi-statically fractured in the three-point-bending (3PB) configuration, while ultrasonic shear waves are periodically generated in the sample at certain time steps $t_\kappa$, $\kappa = \circ, 1, 2, \ldots, 4$. The interaction of probing waves with the propagating damage give rise to transient velocity responses measured on the plate's boundary by a 3D scanning laser Doppler vibrometer. Thus obtained sensory data are then carefully processed to retrieve the associated spectra of scattered displacement fields $\bv^\kappa$ at every $t_\kappa$. On deploying consecutive pairs of multifrequency data $(\bv^\kappa, \bv^{\kappa+1})$, the differential indicators are computed exposing the progress of 3PB-induced damage in the specimen. Verified with in-situ observations, each indicator map successfully reconstructs~(a)~the support of newborn fractures, and~(b)~the loci of discontinuities in the process zone that undergo interfacial evolution in the designated timeframe $[t_\kappa \,\,\, t_{\kappa+1}]$. Further, it is shown that the evolution indicators help better understand the damage mechanism e.g.,~by shining light on the fragmented nature of induced cracks and their coalescence. For completeness, data inversion via reduced and partial-aperture data is investigated, including the one-sided reconstruction. 

\end{abstract}


\end{frontmatter}

\section{Introduction} \label{sec1}

Recent progress in applied mathematics and engineering has led to a suit of robust imaging modalities for real-time sensing in complex environments. State-of-the-art examples include:~ultrasonic surface wave methods~\cite{rose2014}, nonlinear ultrasound~\cite{matl2015}, penetrating-radar techniques~\cite{amin2017}, infrared thermography~\cite{cobl2015}, laser shearography~\cite{hung2013}, X-ray computed tomography~\cite{hank2016}, acoustic tomography imaging~\cite{fuen2019,klos2020,Fatemeh2017}, and deep-leaning schemes~\cite{cant2019,ebra2019,jia2016}. Among these, ultrasonic tomography solutions germane to uncertain and unknown backgrounds are of critical importance as they bear direct relevance to (a) timely detection of degradation in aged, safety-sensitive components~\cite{cobl2015,barn2013}, (b) in-situ monitoring of additive manufacturing processes~\cite{ever2016}, and (c) efficient energy mining from unconventional hydrocarbon and geothermal resources~\citep{Verdon2013b,Verdon2013,Taron2014}. Ongoing efforts in this vein have so far been mostly focused on optimization-based approaches to waveform inversion that typically incur high computational cost at the relevant scales in time and space. Lately, approaches to non-iterative inverse scattering~\cite{cako2016,bonn2019,Fatemeh2017,audi2015,de2018} have been brought under the spotlight for their capabilities pertinent to fast imaging in highly scattering media. While this class of inverse solutions generally demand an a priori characterization of the background for their successful performance, new developments including the differential indicators~\cite{pour2019,audi2015} dispense with this requirement leading to a new class of imaging techniques amenable to environments of uncertain structure and material properties.

This study is focused on the differential imaging functionals~\cite{pour2019} rooted in recent theories on design of sampling methods~\cite{audi2015, Fatemeh2017,Fatemeh2017(2)}. This non-iterative and full-waveform approach is developed for real-time tracking of progressive variations in complex componenets. The idea is to deploy sequential sets of scattered field measurements in the frequency domain to rigorously construct an imaging functional endowed with appropriate invariance with respect to the (unknown) stationary scatterers of the background e.g., pre-existing discontinuities (and inhomogeneities) generated due to imperfect manufacturing or aging. The resulting differential indicators uniquely characterize the support of (geometrically and/or mechanically) evolving process zones in an uncertain domain within a desired timeframe. This is accomplished without the need to reconstruct the entire domain across pertinent scales which may be computationally insurmountable. 

On the verification side, the effectiveness of sampling methods for elastic waveform tomography has been extensively examined by numerical simulations, see e.g.~\cite{cako2016,audi2015,Fatemeh2017,de2018,pour2019}. A systematic experimental investigation of these imaging tools, however, is still lacking. To help bridge the gap, a few recent studies~\cite{baro2016,baro2018} demonstrate successful performance of the classical linear sampling method in a laboratory setting. The present work aims to augment these efforts. In this vein, the differential imaging functionals are deployed in an experimental campaign for spatiotemporal reconstruction of an advancing damage zone in a granite specimen under three-point bending. Prior to fracturing, the sample features a pre-manufactured notch, i.e., a pre-existing scatterer whose support is considered unknown in data inversion.  While fracturing, ultrasonic waves are periodically induced in the specimen at certain sensing steps, and the resulting velocity responses are captured by a 3D scanning laser Doppler vibrometer over the plate's edges. Such sensory data are then carefully transformed into the frequency domain, and used to recover the support of evolution in a sequence. Here, the inverse solution is adapted to test data and reformulated for multi-frequency reconstruction. It is shown that the differential indicators expose not only the process zone's geometry, but also the support of elastically evolving interfaces. The latter is proven to be immediately relevant to damage propagation in the future timeframes. The influences of key testing parameters on the fidelity of reconstruction -- including the source and measurement aperture, and sensing resolution, are investigated using experimental data.

This paper is organized as follows. \textcolor{black}{Section}~\ref{prelim} formulates the direct scattering problem within the context of laboratory experiments, and provides an overview of the data inversion platform. \textcolor{black}{Section}~\ref{exp_set} describes the experimental procedure and showcases the ``raw" measurements. \textcolor{black}{Section}~\ref{DSP} includes a detailed account of signal processing in time and space in preparation for data inversion. \textcolor{black}{Section}~\ref{DI} computes the differential imaging functionals using multi-frequency data. \textcolor{black}{Section}~\ref{RE} presents and discusses the results.

\section{Theoretical foundation} \label{prelim}
   
To provide a framework for the ensuing experimental campaign, this section briefly delineates the theory of differential imaging~\cite{pour2019}.       

\vspace{-1mm}
\subsection{Problem statement} \lb{FP}

Let $\mathcal{B} \subset \R^d$, $d = 2, 3$, denote a finite elastic body characterized by mass density $\rho$, and Lam\'{e} parameters $\mu$ and~$\lambda$, which henceforth is referred to as the \emph{baseline model}. Two sets of \emph{unknown} scatterers are embedded in $\mathcal{B}$, namely:~(i)~a \emph{time-invariant} network of pre-existing interfaces $\Gamma_{\! \circ}$ which includes manufacturing-induced dislocations and (micro) cracks, and~(ii)~an \emph{evolving} set of discontinuities $\Gamma(t)$ driven by various chemo-physical reactions in operational environments. At time $t$, the support of scatterers $\Gamma_{\! \circ} \cup \Gamma(t)$ is possibly disjoint, of arbitrary shape, and may be decomposed into $N_t$ smooth open subsets $\Gamma_n$. The support of every $\Gamma_n$ may be arbitrarily extended to a closed Lipschitz surface $\partial \text{\sf D}_n$ enclosing a bounded simply connected domain $\text{\sf D}_n \subset \R^d$, so that $\Gamma_{\!\circ} \cup \Gamma(t) \!=\! {\textstyle \bigcup_{n =1}^{N_t}} \Gamma_n \subset {\textstyle \bigcup_{n =1}^{N_t}}  \partial \text{\sf D}_n$. The contact condition at the surface of $\Gamma_{\!\circ}$ (\emph{resp.}~$\Gamma(t)$) is discontinuous characterized by a symmetric and heterogeneous interfacial stiffness matrix $\bK_{\!\circ}(\bxi), \, \bxi \in \Gamma_{\!\circ}$ (\emph{resp.}~$\bK(\bxi,t), \, \bxi \in \Gamma(t)$) synthesizing the spatially varying nature of rough and/or multiphasic interfaces. Here, $\bK_{\!\circ}$ and $\bK$ are a priori unknown. However, it is assumed that energy dissipation at interfaces remains negligible during ultrasonic measurements. 

The domain $\mathcal{B}$ is subject to periodic ultrasonic inspections at time steps $t_\kappa = \lbrace t_1, t_2, ... \rbrace$. At every $t_\kappa$, the specimen is excited by an ultrasonic source on its external boundary $\partial \mathcal{B}$ so that the corresponding incident field $\bu^{\textrm{f}}(\bxi,t)$ in the baseline model is governed by 
\vspace{-1.5 mm}
\beq\lb{uf}
\begin{aligned}
&\nabla \exs\sip\exs [\bC \exs \colon \! \nabla \bu^{\textrm{f}}\exs](\bxi,t) \,-\, \rho \exs \ddot{\bu}^{\textrm{f}}(\bxi,t) ~=~ \bzero, \quad &  \big(\bxi \in {\mathcal{B}}, t \in (0,T] \big) \\*[0.5mm]
&\bn \exs\sip\exs \bC \exs \colon \!  \nabla  \bu^{\textrm{f}}(\bxi,t)~=~\bg(\bxi,t),  \quad & \big(\bxi \in \partial {\mathcal{B}}_t, t \in (0,T] \big) \\*[0.5mm]
& \bu^{\textrm{f}}(\bxi,t)~=~\bzero,   \quad & \big(\bxi \in \partial\mathcal{B}_u, t \in (0,T] \big)\\*[0.5mm]
& \bu^{\textrm{f}}(\bxi,0)~=~ \dot{\bu}^{\textrm{f}}(\bxi,0)~=~ \bzero,   \quad & \big(\bxi \in \overline{\mathcal{B}}, t = 0 \big)
\end{aligned}   
\vspace{-1.5 mm}
\eeq  
where the fourth-order elasticity tensor $\bC = \lambda\bI_2\!\otimes\!\bI_2 + 2\mu\bI_4$ with $\bI_m \,(m\!=\!2,4)$ denoting the $m$th-order symmetric identity tensor; the single and double over-dots indicate first- and second- order time derivates, respectively; $T$ signifies the testing interval; $\bn$ is the unit outward normal to the sample's boundary $\partial \mathcal{B}$; $\bg(\bxi,t)$ represents the external traction on the Neumann part of the boundary $\partial {\mathcal{B}}_t \subset \partial {\mathcal{B}}$ which includes the source input; the displacement vanishes on the boundary's Dirichlet part $\partial\mathcal{B}_u \subset \partial\mathcal{B}$; and, overline indicates the closure of a set e.g.,~$\overline{\mathcal{B}} = {\mathcal{B}} \cup \partial {\mathcal{B}}$. At every sensing step $t_\kappa$, the interaction of $\bu^{\textrm{f}}$ with the hidden scatterers $\Gamma_{\! \circ} \cup \Gamma(t_\kappa)$ in the specimen gives rise to the total field $\bu^\kappa(\bxi,t)$ satisfying      
\vspace{-1.5 mm}
\beq\lb{uk}
\begin{aligned}
&\nabla \exs\sip\exs [\bC \exs \colon \! \nabla \bu^\kappa\exs](\bxi,t) \,-\, \rho \exs \ddot{\bu}^\kappa(\bxi,t) ~=~ \bzero, \quad &  \big(\bxi \in {\mathcal{B}}\backslash\lbrace \Gamma_{\!\circ} \cup \Gamma_\kappa \rbrace, t \in (0,T] \big) \\*[0.5mm]
&\bn_\alpha \sip\exs \bC \exs \colon \!  \nabla  \bu^\kappa(\bxi,t)~=~\bK_{\!\alpha}(\bxi) \llbracket \bu^\kappa \rrbracket(\bxi,t),  \quad & \big(\bxi \in \Gamma_{\!\circ}\cup\Gamma_\kappa, t \in (0,T] \big) \\*[0.5mm]
&\bn \exs\sip\exs \bC \exs \colon \!  \nabla  \bu^\kappa(\bxi,t)~=~\bg(\bxi,t),  \quad & \big(\bxi \in \partial {\mathcal{B}}_t, t \in (0,T] \big) \\*[0.5mm]
& \bu^\kappa(\bxi,t)~=~\bzero,   \quad & \big(\bxi \in \partial\mathcal{B}_u, t \in (0,T] \big)\\*[0.5mm]
& \bu^\kappa(\bxi,0)~=~ \dot{\bu}^\kappa(\bxi,0)~=~ \bzero,   \quad & \big(\bxi \in \overline{\mathcal{B}}, t = 0 \big)
\end{aligned}   
\vspace{-0.5 mm}
\eeq  
where $\Gamma_\kappa \nxs:= \Gamma(t_\kappa)$; $\llbracket \bu^\kappa \rrbracket(\bxi,t)$ indicates the jump in displacement field across $\bxi \in \Gamma_\kappa\cup\Gamma_{\!\circ}$;   
\vspace{-1.5 mm}
\beq\label{nK}\nonumber
\bK_{\!\alpha}(\bxi) \,=\,
\!\left\{\begin{array}{l}
\!\! \bK_\circ(\bxi), \qquad\,\,\,\, \bxi \in \Gamma_{\!\circ}\backslash\tilde{\Gamma}_{\!\circ} \!\!\! \\*[0.5mm]
\!\! \bK(\bxi,t_\kappa), \quad\,\,\,\,\, \bxi \in \Gamma_\kappa\nxs\cup\tilde{\Gamma}_{\!\circ}  \!\!\!
\end{array}\right., \qquad 
\bn_\alpha(\bxi) \,=\,
\!\left\{\begin{array}{l}
\!\! \bn_\circ(\bxi), \qquad \bxi \in \Gamma_{\!\circ} \!\!\! \\*[0.5mm]
\!\! \bn_\kappa(\bxi), \qquad \bxi \in \Gamma_\kappa  \!\!\!
\end{array}\right.,
\vspace{-1.5 mm}
\eeq
wherein
$\overline{\tilde{\Gamma}_{\!\circ}} \, \colon\!\!\!= \overline{\big\lbrace \bxi \subset {\Gamma}_{\!\circ}\!: \,\,\, \bK(\bxi,t_\kappa) \neq \bK_{\!\circ}(\bxi) \big\rbrace},$ 
signifying a subset of $\Gamma_{\!\circ}$ which undergoes interfacial evolution between $[t_1 \,\, t_\kappa]$; and, $\bn_\circ$ (\emph{resp.}~$\bn_\kappa$) indicates the unit normal vector on $\Gamma_{\!\circ}$ (\emph{resp.}~$\Gamma_\kappa$) which on recalling $\Gamma_{\!\circ} \cup \Gamma_\kappa \subset {\textstyle \bigcup_{n =1}^{N_t}}  \partial \text{\sf D}_n$, is outward to $\text{\sf D}_n$. Such induced wave motion is then measured over the observation surface $S\obs\subset \partial {\mathcal{B}}_t$. In this setting, the periodic experiments furnish a sequential set of sensory data $\bu^\kappa$ on $S\obs$ associated with ultrasonic excitations on the incident surface $S^{\textrm{inc}\!}\! \subset \partial {\mathcal{B}}_t$. Note that the corresponding scattered displacement fields $\bv^\kappa\!$ may be computed as the following,
\vspace{-1.5 mm}
\beq\lb{SDF}
\bv^\kappa(\bxi,t) = [\bu^\kappa- \bu^{\textrm{f}}](\bxi,t), \qquad  \kappa = 1,2, \ldots, \qquad \bxi \in S\obs, \, t \in (0,T].
\vspace{-1.5 mm}
\eeq  

To assist the inverse analysis, let us introduce the relevant function spaces as the following,
\vspace*{-1.5mm}
\beq\lb{funS2}
\begin{aligned}
&H^{\pm \frac{1}{2}}(\Gamma_{\!\circ} \cup \Gamma_\kappa) ~:=~\big\lbrace f\big|_{\Gamma_{\!\circ} \cup \Gamma_\kappa} \! \colon \,\,\, f \in H^{\pm \frac{1}{2}}(\partial \text{\sf D}_t) \big\rbrace, \\*[0.0 mm]
& \tilde{H}^{\pm \frac{1}{2}}(\Gamma_{\!\circ} \cup \Gamma_\kappa) ~:=~\big\lbrace  f \in H^{\pm\frac{1}{2}}(\partial \text{\sf D}_t) \colon  \,\,\, \text{supp}(f) \subset \overline{\Gamma_{\!\circ} \cup \Gamma_\kappa} \exs \big\rbrace,
\end{aligned}
\vspace*{-1.5mm}
\eeq
where $\text{\sf D}_t = {\textstyle \bigcup_{n =1}^{N_t}} \text{\sf D}_n$ is a multiply connected Lipschitz domain of bounded support such that $\Gamma_{\!\circ} \cup \Gamma_\kappa \subset \partial \text{\sf D}_t$, and $\overline{\Gamma_{\!\circ} \cup \Gamma_\kappa} \colon \!\!\! =  (\Gamma_{\!\circ} \cup \Gamma_\kappa ) \cup ( \partial\Gamma_{\!\circ} \cup \partial\Gamma_\kappa )$ denotes the closure of $\Gamma_{\!\circ} \cup \Gamma_\kappa \!=\! {\textstyle \bigcup_{n =1}^{N}} \Gamma_n$. Recall that every $\Gamma_n$ is an open set (relative to $\partial \text{\sf D}_n$) with a positive surface measure. It should be mentioned that since $\bv^\kappa \in  H^1({\mathcal{B}}\backslash\lbrace\Gamma_{\!\circ} \cup \Gamma_\kappa \rbrace)^3$, then by trace theorems,  $\llbracket \bv^\kappa \rrbracket\in\tilde{H}^{1/2}(\Gamma_{\!\circ} \cup \Gamma_\kappa)^3$.

\vspace{-1mm}
\subsection{Inverse solution} \lb{InvS}

\emph{Differential imaging functionals} deploy consecutive pairs of scattered field measurements $(\bv^{\kappa}, \bv^{\kappa+1})$ to reconstruct the support of (geometric and mechanical) evolution $\hat\Gamma_{\kappa+1}\nxs \cup \tilde\Gamma_{\kappa+1}\!$, in the associated timeframe $[t_\kappa \,\,\, t_{\kappa+1}]$. This is accomplished without the need to recover all the pre-existing scatterers $\Gamma_{\!\circ}\nxs \cup \Gamma_{\!\kappa}$ at $t_\kappa$. The evolution support consists of two subsets, namely:~(i)~newborn elastic interfaces
\vspace*{-1.5mm}
\beq \lb{HG}
\hat{\Gamma}_{\kappa+1} \, \colon\!\!\!= {\Gamma}_{\kappa+1} \backslash \overline{{\Gamma}_{\kappa}}, \qquad \kappa = 1,2,3,
\vspace*{-1.5mm}
\eeq   
and~(ii)~interfacially modified contacts $\tilde{\Gamma}_{\kappa+1}$,
\vspace*{-1.5mm}
\beq \lb{TG}
\overline{\tilde{\Gamma}_{\kappa+1}} \, \colon\!\!\!= \overline{\big\lbrace \bxi \subset {\Gamma}_{\!\kappa} \cup {\Gamma}_{\!\circ}\!: \,\,\, \bK(\bxi,t_{\kappa}) \neq \bK(\bxi,t_{\kappa+1}) \big\rbrace}, \qquad \kappa = 1,2,3.
\vspace*{-1.5mm}
\eeq   

Targeted imaging of $\hat\Gamma_{\kappa+1}\! \cup \tilde\Gamma_{\kappa+1}\!$ is conducted in the frequency domain via synthetic wavefront shaping, followed by invoking functionals of systematic invariance with respect to the stationary scatterers $\Gamma_{\!\circ}\nxs \cup \Gamma_{\!\kappa}$. At every $t_\kappa$, the spectrum of scattered displacement fields $\bv^{\kappa}$ on $S\obs$ over the bandwidth $\Omega := [\omega_{\min}\,\,\, \omega_{\max}] \subset \R^+$ is used to \emph{non-iteratively} compute the associated wavefront densities $\bg$ on $S^{\textrm{inc}}\!$. To this end, the scattering operator $\Lambda_\kappa\!:\, L^2(S^{\text{inc}})^3\nxs\times L^2(\Omega)^3  \exs\to\exs L^2(S\obs)^3\nxs\times L^2(\Omega)^3$ is constructed on the basis of test data as follows,
\vspace{-1.5mm}
\beq\lb{So} 
\Lambda_\kappa(\bg)(\bxi,\omega) ~=\,  \int_{S^{\text{inc}\!}} \bV^{\kappa}(\bxi,\by;\omega) \sip \bg(\by,\omega) \,\, \text{d}S_{\by}, \qquad \bg \in L^2(S^{\text{inc}})^3\nxs\times L^2(\Omega)^3, \quad\!\! \bxi \in S\obs, \,\,\, \omega \in \Omega.
\vspace{-1.5mm}
\eeq 
On denoting by $F(\cdot)$ the Fourier transform operator, $V^{\kappa}_{ij}(\bxi,\by;\omega)$, $i,j\!=\!1,2,3$, in~\eqref{So} indicates the $i^{\textrm{th}}$ component of the Fourier transformed displacement $F(\bv^{\kappa})(\bxi,\omega) \in L^2(S\obs)^3\nxs\times L^2(\Omega)^3$ measured at $\bxi \in S\obs$ with frequency $\omega \in \Omega$ due to excitation at $\by \in S^{\textrm{inc}}$ in the $j^{\textrm{th}}$ direction. Recall that $\kappa$ signifies the sensing step.

On the other hand, let us consider the search volume $\mathcal{S} \subset \mathcal{B} \subset \R^{d}$ in the \emph{baseline model}, and define a set of trial dislocations $L(\bx_\circ, \bR) \subset \mathcal{S}$ such that for every pair $(\bx_\circ,\bR)$, $L\colon\!\!=\bx_\circ\!+\bR{\sf L}$ specifies a smooth arbitrary-shaped fracture~$\sf L$~at $\bx_\circ \subset \mathcal{S}$ whose orientation is identified by a unitary rotation matrix $\bR\!\in\!U(3)$. In this setting, the scattering pattern $\bPhi_L \colon \tilde{H}^{1/2}(L)^3 \nxs\times L^2(\Omega)^3 \rightarrow L^2(S\obs)^3 \nxs\times L^2(\Omega)^3$ on $S\obs$ -- generated by $L(\bx_\circ, \bR)$, as a sole scatterer in $\mathcal{B}$, endowed with an admissible displacement density $\ba(\bxi,\omega)\!\in\!\tilde{H}^{1/2}(L)^3 \nxs\times L^2(\Omega)^3$ is governed by 
\vspace{-1.5mm}
\beq\lb{PhiL}
\begin{aligned}
&\nabla \nxs\cdot [\bC \exs \colon \! \nabla \bPhi_L](\bxi,\omega) \,+\, \rho \omega^2\bPhi_L(\bxi,\omega)~=~\bzero, \quad & \big(\bxi \in {\mathcal{B}}\backslash L, \omega \in \Omega \big) \\*[0.5mm]
&\bn \nxs\cdot \bC \exs \colon \!  \nabla  \bPhi_L(\bxi,\omega)~=~\bzero,  \quad & \big(\bxi \in \partial{\mathcal{B}}_t, \omega \in \Omega \big) \\*[0.5mm]
&\bPhi_L(\bxi,\omega)~=~\bzero,  \quad & \big(\bxi \in \partial{\mathcal{B}}_u, \omega \in \Omega \big) \\*[0.5mm]
& \llbracket \bPhi_L \rrbracket(\bxi,\omega)~=~\boldsymbol{a}(\bxi,\omega). \quad & \big(\bxi \in L, \omega \in \Omega \big)
\end{aligned}     
\vspace{-1.5mm}
\eeq
Given~\eqref{PhiL}, one may generate a library of physically-consistent scattering patterns on $S\obs$ for a grid of trial pairs $(\bx_\circ,\bR)$ sampling $\mathcal{S}\!\times\! U(3)$. 

The underpinning concept of wavefront shaping is that when the trial dislocation $L \subset \Gamma_{\!\circ}\nxs \cup \Gamma_{\!\kappa}$, the pattern $\bPhi_L \in L^2(S\obs)^3 \nxs\times L^2(\Omega)^3$ may be recovered from experimental data by probing the range of operator $\Lambda_\kappa$ through solving 
\vspace{-1.5 mm}
\beq\lb{FF}
\Lambda_\kappa\exs \bg^\kappa~\simeq~\bPhi_L, 
\vspace{-1.5 mm}
\eeq
for the wavefront densities $\bg^\kappa(\bxi, \omega)$ on $\bxi \in S^{\textrm{inc}}$ for every frequency $\omega \in \Omega$.
Based on~\eqref{FF}, the principal theorems of differential imaging~\cite[Theorems 4.3, 4.5, 4.8]{pour2019} rigorously establish the distinct behavior of the solution $\bg^\kappa$ in terms of $L$, particularly when $L \subset \Gamma_{\!\circ}\nxs \cup \Gamma_{\!\kappa}$. Owing to the ill-posed nature of~\eqref{FF}, first,~\cite[Theorems 4.3]{pour2019} furnishes a carefully designed approximate solution to~\eqref{FF} through minimizing the regularized cost functional
\vspace{-1.5mm}
\beq\lb{GCfn}
\mathfrak{J}_\kappa(\bg;\bPhi_L,\omega) \,:=\,\,\exs  \norms{\nxs \Lambda_{\kappa}\exs\bg\,-\,\bPhi_L \nxs}_{L^2}^2 \!\,+\exs\,\exs\gamma\, \big(\bg, \Upsilon_\kappa\exs\bg\big)_{L^2} \!\,+\exs\, \gamma^{1-\chi}\exs\delta  \norms{\nxs \bg \nxs}^2_{L^2}, \qquad  \Upsilon_\kappa =  \big( \Lambda_{\kappa}^* \Lambda_{\kappa} \nxs\big)^{{1}/{2}},
\vspace{-1.5mm}
\eeq
where $\chi \in \, ]0, 1[$ is a constant independent of $\bg$; $\delta>0$ stands for a measure of noise in data; $\gamma>0$ represents the regularization parameter; and, $\Lambda_{\kappa}^*$ is the adjoint of $\Lambda_{\kappa}$. It is further shown that $\mathfrak{J}_\kappa$ is convex and its minimizer $\bg^\kappa(\bxi, \omega)$ may be obtained non-iteratively according to section~\ref{DI}. 

\vspace{-2.5mm}
\begin{rem}[on $\Upsilon_\kappa$]\label{ups}
It should be mentioned that the operator $\Upsilon_\kappa(\Lambda_\kappa)$ in~\eqref{GCfn} replaces 
\vspace{-1.5mm}
\beq\lb{Fsd}
\Lambda_{\kappa_\sharp}\,\colon \!\!\!=\, \frac{1}{2} \big{|}\Lambda_{\kappa}+\Lambda_{\kappa}^*\big{|} \:+\: \big{|}\frac{1}{2 \textrm{\emph{i}}} (\Lambda_{\kappa}\nxs-\Lambda_{\kappa}^*)\big{|}, 
\vspace{-1.5mm}
\eeq
in~\cite{pour2019} due to a particular implication of the latter that the discretized operator $\Lambda_{\kappa}$ must be a square matrix, i.e.,~the number of ultrasonic sources should equal the number of observation points. As evidenced in section~\ref{exp_set}, this may not be plausible or efficient in practice. The operator $\Upsilon_\kappa$ deployed in~\eqref{GCfn} relaxes this constraint, while still carrying the fundamental properties required by the theorems of differential imaging. The latter holds provided that the system's energy dissipation may be assumed negligible during the testing period $(0, T]$ and that the operator $\Lambda_{\kappa}$ is normal~\cite{Kirsch2008}. As a result, given the factorization 
\vspace{-1.5mm}
\beq\lb{facts1}
\Lambda_{\kappa} ~=~ \mathcal{H}_{\kappa}^* \exs T_{\kappa} \exs \mathcal{H}_{\kappa}, 
\vspace{-1.5mm}
\eeq
according to~\cite[Remark 3.3]{pour2019} with a coercive middle operator $T_{\kappa}$, Theorem 1.23 of~\cite{Kirsch2008} reads that there exists a second factorization 
\vspace{-1.5mm}
\beq\lb{facts2}
\Lambda_{\kappa} ~=~ \big( \Lambda_{\kappa}^* \Lambda_{\kappa} \nxs\big)^{\!\frac{1}{4}} \, \text{\sf T}_{\kappa} \exs \big( \Lambda_{\kappa}^* \Lambda_{\kappa} \nxs\big)^{\!\frac{1}{4}}, 
\vspace{-1.5mm}
\eeq 
such that $\text{\sf T}_{\kappa}$ is coercive, and thus, the ranges of $\mathcal{H}_{\kappa}^*$ and $\big( \Lambda_{\kappa}^* \Lambda_{\kappa} \nxs\big)^{\!\frac{1}{4}}$ coincide, warranting the use of $\Upsilon_\kappa$ in~\eqref{GCfn}.       
\vspace{-2.5mm}
\end{rem}

In light of remark~\ref{ups} and~\cite[Theorems 4.3]{pour2019}, one may observe that as $\gamma \rightarrow 0$, the solution $\bg^\kappa\!$ to~\eqref{FF} remains bounded if and only if $L \subset \Gamma_{\!\circ}\nxs \cup \Gamma_{\!\kappa}$. More specifically,~at every $t_\kappa$,
\vspace{-1.5mm}
\beq\lb{statG1}
\begin{aligned}
& \text{if}\,\,\, L \subset \Gamma_{\!\circ}\nxs \cup \Gamma_{\!\kappa} ~\Rightarrow~ \limsup\limits_{\gamma \rightarrow 0}\limsup\limits_{\delta \rightarrow 0} \Big( \nxs (\exs \bg^{\kappa}, \Upsilon_\kappa\exs\bg^{\kappa})_{L^2} \!\,+\, \gamma^{-\chi}\exs\delta  \norms{\nxs \bg^{\kappa} \nxs}^2_{L^2}  \!\!\Big) \,<\, \infty, \\
&\text{if}\,\,\, L \not\subset \Gamma_{\!\circ}\nxs \cup \Gamma_{\!\kappa} ~\Rightarrow~ \liminf\limits_{\gamma \rightarrow 0}\liminf\limits_{\delta \rightarrow 0} \Big( \nxs (\exs \bg^{\kappa}, \Upsilon_\kappa\exs\bg^{\kappa})_{L^2} \!\,+\, \gamma^{-\chi}\exs\delta  \norms{\nxs \bg^{\kappa} \nxs}^2_{L^2}  \!\!\Big) \,=\, \infty.
 \end{aligned}
\vspace{-1.5mm}
\eeq

Also, given $\bg^\kappa\!$ minimizing $\mathfrak{J}_\kappa$ at every $t_\kappa$, from Theorems 4.5 and 4.8 of~\cite{pour2019}, one may show that the functional 
\vspace{-1.5mm}
\beq \lb{Inv2}
\mathfrak{I}_{\kappa}(\bg^\kappa, \bg^{\kappa+1};\omega) := \big(\, \bg^{\kappa+1}-\,\bg^\kappa,\, \Upsilon_\kappa(\bg^{\kappa+1}-\,\bg^\kappa)\big)_{L^2} + \exs\delta \norms{\bg^{\kappa+1}-\,\bg^\kappa\!}^2_{L^2}, 
\vspace{-1.5mm}
\eeq
remains invariant at the loci of stationary scatterers ${\Gamma}_{\!\circ} \cup \Gamma_{\!\kappa} \backslash \tilde{\Gamma}_{\kappa+1}\!$ for all $t_\kappa$. More specifically, it may be shown that 
\vspace{-1.5mm}
\beq\lb{statG2}
\begin{aligned}
& \text{if}\,\,\, L\subset\Gamma_{\!\kappa} \cup {\Gamma}_{\!\circ} \backslash \tilde{\Gamma}_{\kappa+1} \,\,\,\,\,\, ~\Rightarrow~ \!\!\!\!\!\! &  \lim\limits_{\gamma \rightarrow 0}\liminf\limits_{\delta \rightarrow 0}\mathfrak{I}_\kappa(\,\bg^\kappa, \bg^{\kappa+1};\omega) \,=\, 0, \\
& \text{if}\,\,\, L\subset\tilde{\Gamma}_{\kappa+1}  \,\,\,\,\,\,\,\,\, ~\Rightarrow~ \!\!\!\!\!\!\!\!\!\!\!\!  & 0 < \lim\limits_{\gamma \rightarrow 0}\liminf\limits_{\delta \rightarrow 0}\mathfrak{I}_\kappa(\,\bg^\kappa, \bg^{\kappa+1};\omega) < \infty,\\
& \text{if}\,\,\, L\subset\hat{\Gamma}_{\!\kappa+1} \,\,\,\,\,\,\,\,\, ~\Rightarrow~ \!\!\!\!\!\!\!\!\!\!\!\!  & \lim\limits_{\gamma \rightarrow 0}\liminf\limits_{\delta \rightarrow 0}\mathfrak{I}_\kappa(\,\bg^\kappa, \bg^{\kappa+1};\omega) \,=\, \infty.
 \end{aligned}
\vspace{-1.5mm}
\eeq

In light of~\eqref{statG1} and~\eqref{statG2}, the evolution indicator functionals are defined by  
\vspace{-1.5mm}
\beq\lb{EIFn0}
\begin{aligned}
& \mathcal{D}_\kappa(\bg^\kappa, \bg^{\kappa+1};\omega) := \frac{1}{\sqrt{\mathfrak{I}_{\kappa+1}(\bzero, \bg^{\kappa+1};\omega) \big{[}1+ \mathfrak{I}_{\kappa+1}(\bzero, \bg^{\kappa+1};\omega)\exs {\mathfrak{I}_{\kappa}(\bg^\kappa, \bg^{\kappa+1};\omega)}^{-1}\big{]}}},  \\
&\tilde{\mathcal{D}}_\kappa(\bg^\kappa, \bg^{\kappa+1};\omega)  := \frac{1}{\sqrt{\mathfrak{I}_{\kappa}(\bg^{\kappa}, \bzero;\omega) + \mathfrak{I}_{\kappa+1}(\bzero, \bg^{\kappa+1};\omega) \big{[}1+\mathfrak{I}_{\kappa}(\bg^{\kappa}, \bzero;\omega)\exs {\mathfrak{I}_{\kappa}(\bg^\kappa, \bg^{\kappa+1};\omega)}^{-1}\big{]}}}.
\end{aligned}
\vspace*{-1.5mm}
\eeq
Here, $\tilde{\mathcal{D}}_{\kappa}$ illuminates the support of mechanical evolution within $[t_{\kappa} \,\,\, t_{\kappa+1}]$ by achieving its highest values at the loci of interfacially modified contacts $\tilde{\Gamma}_{\kappa+1}$ according to~\eqref{TG}. On the other hand, $\mathcal{D}_\kappa$ reconstructs the support of evolution more holistically both in terms of the newly born interfaces $\hat{\Gamma}_{\kappa+1}$ i.e., the geometric evolution as in~\eqref{HG}, as well as the elastically modified contacts $\tilde{\Gamma}_{\kappa+1}$. More rigorously, the behavior of $\mathcal{D}_\kappa$ and $\tilde{\mathcal{D}}_\kappa$ within the search volume $\mathcal{S}\subset\mathcal{B}$ may be characterized as the following, 
\vspace{-1.5mm}
\beq\lb{EIFn2}
\begin{aligned}
& \text{if}\,\,\, L \subset \tilde{\Gamma}_{\kappa+1} \cup \hat{\Gamma}_{\kappa+1} \quad \iff \quad  \lim\limits_{\gamma \rightarrow 0} \exs \liminf\limits_{\delta \rightarrow 0} \exs \mathcal{D}_\kappa(\bg^\kappa, \bg^{\kappa+1};\omega) \,>\, 0, \\
& \text{if}\,\,\, L \subset \mathcal{S}\backslash \lbrace\tilde{\Gamma}_{\kappa+1} \cup \hat{\Gamma}_{\kappa+1} \nxs \rbrace \quad \iff \quad \liminf\limits_{\gamma \rightarrow 0} \exs \liminf\limits_{\delta \rightarrow 0} \exs \mathcal{D}_\kappa(\bg^\kappa, \bg^{\kappa+1};\omega) = 0, \\
& \text{if}\,\,\, L \subset  \tilde{\Gamma}_{\kappa+1} \quad \iff \quad \lim\limits_{\gamma \rightarrow 0} \exs \liminf\limits_{\delta \rightarrow 0} \exs \tilde{\mathcal{D}}_\kappa(\bg^\kappa, \bg^{\kappa+1};\omega) \,>\, 0, \\
& \text{if}\,\,\, L \subset \mathcal{S}\backslash \tilde{\Gamma}_{\kappa+1} \quad \iff \quad   \liminf\limits_{\gamma \rightarrow 0} \exs \liminf\limits_{\delta \rightarrow 0} \exs \tilde{\mathcal{D}}_\kappa(\bg^\kappa, \bg^{\kappa+1};\omega) = 0.
\end{aligned}
\vspace*{-1.5mm}
\eeq

\vspace*{-3mm}
\section{Experimental campaign} \label{exp_set}

\noindent Experiments are performed on a prismatic specimen of charcoal granite of dimensions $0.96$m $\!\!\times\exs 0.3$m $\!\!\times\exs 0.03$m, mass density $\rho\!=\!2750$kg/m$^3$, nominal Poisson's ratio $\nu\!=\!0.23$, and nominal Young's modulus $E\!=\!62.6$GPa. These values are identified via a uniaxial compression test on a cylindrical sample of the same material. 

The testing procedure is twofold involving (i) quasi-static fracturing of the specimen, and (ii) periodic ultrasonic excitation and sensing.

\begin{figure}[!tp] \vspace*{-0mm}
\center\includegraphics[width=0.91\linewidth]{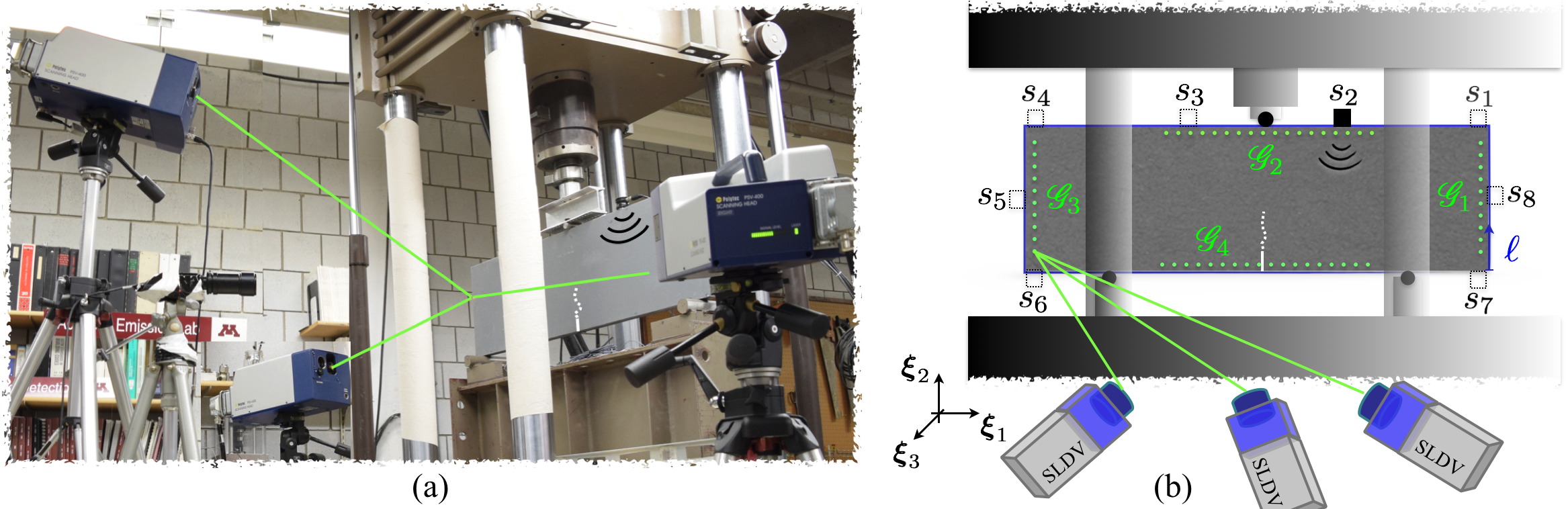} \vspace{-3.5mm}
\caption{Testing set-up for differential tomography of fracture evolution:~(a)~a prismatic slab of charcoal granite is quasi-statically fractured by a closed-loop, servo-hydraulic $1000$kN MTS load frame in the three-point-bending (3PB) configuration with the crack mouth opening displacement (CMOD) as the feedback signal; the CMOD is held constant at approximately $90\%$, $75\%$, and $60\%$ of the maximum load in the post-peak regime for ultrasonic testing;~(b)~shear waves are generated periodically by a piezoelectric source at~$s_i$ \mbox{$(i\!=\!1,2,\ldots,8)$}; the triaxial particle velocity field is then captured by~SLDV over the designated scanning grid~$\bigcup_{i=1}^4\mathcal{G}_i$.} \label{Exp-sch} \vspace{-4mm}
\end{figure}

A notch of length $4$cm and width $1.5$mm is manufactured at the bottom center of specimen. The sample is then fractured in the three-point-bending (3PB) configuration as shown in Fig.~\ref{Exp-sch} by a closed-loop, servo-hydraulic, $1000$kN MTS load frame such that the crack propagation is controlled by the crack mouth opening displacement (CMOD) measured by a clip gage. The loading process is monotonic with respect to the CMOD at a constant rate of $0.1\mu$m/s. However, at $275\mu$m, $300\mu$m, and $325\mu$m -- corresponding respectively to nearly $90\%$, $75\%$, and $60\%$ of the maximum load in the post-peak regime, the CMOD is held constant for a period of 4-6 hours for ultrasonic probing. 

Ultrasonic experiments are conducted at five time steps $t =  \lbrace t_\circ,t_1, t_2, \ldots, t_4 \rbrace$. At $t_\circ$, prior to notching, the granite slab is intact mounted on the load frame without prestressing. Waveforms measured at this step furnish the ``baseline'' response of the system associated with the incident field $\bu\ff(\bxi,t)$. This is required for computing the scattered field $\bv^\kappa(\bxi,t) := \bu^\kappa(\bxi,t) - \bu\ff(\bxi,t)$ at any future sensing step $t_\kappa$, $\kappa = 1, 2, 3, 4$, wherein $\bu^\kappa(\bxi,t)$ represents the \emph{total} field measurements at $t_\kappa$. Bear in mind that the differential indicators deploy the spectrum of \emph{scattered} field $\bv^\kappa$ for data inversion. Experiments are periodically repeated after notching at $t = t_1, \ldots, t_4$ in a similar setting in terms of the specimen configuration, transducer locations, illuminating wavelet, and scanning area. At $t_1$, prior to fracturing, the prestress remains zero. At $t_2$, $t_3$, and $t_4$, however, fracturing is underway and the applied force by the load frame may be estimated respectively by $12.7$kN, $10.5$kN, and $8.5$kN in the post-peak regime.
  
Every sensing step $t_\kappa$, $\kappa = \circ, 1, \ldots, 4$ entails eight ultrasonic experiments where the sample is excited by an \emph{in-plane} shear wave from one of the designated source locations $s_1, s_2, \ldots, s_{8}$ shown in Fig.~\ref{Exp-sch}(b). Shear waves are generated by a 0.5 MHz piezoelectric transducer (V151-RB by Olympus, Inc.)~whose diameter of 32 mm is almost commensurate with the granite thickness. The transducer is aligned with the granite mid-plane along $\bxi_3$ minimizing the out-of-plane excitation. The incident signal is a five-cycle burst of the form 
\begin{equation}\label{wavelet}
H({\sf f_c}t) \, H(5\!-\!{\sf f_c}t) \, \sin\big(0.2 \pi {\sf f_c} t\big) \, \sin\big(2 \pi {\sf f_c} t\big), 
\end{equation}
where ${\sf f_c}\!=\!30\mbox{kHz}$ denotes the center frequency, and $H$ is the Heaviside step function. The induced wave motion from each source location is measured by a 3D Scanning Laser Doppler Vibrometer (SLDV) as shown in Fig.~\ref{Exp-sch}(a). The PSV-400-3D SLDV system by Polytec, Inc.~is capable of capturing the triaxial components of particle velocity on the surface of solids over a designated scanning grid. Its measurement (\emph{resp.}~spatial) resolution is better than 1$\mu${m}/s (\emph{resp.}~0.1mm) within the frequency range DC-1MHz, facilitating waveform sensing in the nanometer scale in terms of displacement~\citep{Polytec}.  

\vspace{-2.5mm}
\begin{rem}[nature of the wave motion]\label{pw}
Measurements may be interpreted in the context of \emph{plane stress} approximation -- related to the elastic analysis of thin plates~\citep{Mal1969}, whereby the particle motion is considered invariant through the thickness of specimen. In this setting, the \emph{effective} Poisson's ratio and Young's modulus are respectively identified by $\nu'\!=\!\nu/(1+\nu)$ and $E'\!=\!E(1-\nu'^2)$~\citep{Mal1969}, resulting in the shear (S-) and compressional (P-) wave velocities 
\vspace{-1.75mm}
\begin{equation}\label{cps}
c_s ~=~ \sqrt{\frac{E}{2(1+\nu)\rho}} ~=~ 3041 \, \, \mbox{m/s}, \qquad  c_p ~=~ \sqrt{\frac{E}{(1-\nu^2)\rho}} ~=~ 4901 \, \, \mbox{m/s}. 
\vspace{-1.25mm}
\end{equation}
Observe that the shear wavelength $\lambda_s$ in the specimen may be approximated by $10$cm at $30\mbox{kHz}$, giving the shear-wavelenghth-to-plate-thickness ratio of $\lambda_s/h\nxs\gtrsim\nxs3.3$. In this range, the phase error committed by the plane stress approximation is about $3\%$~\citep{Lamb1917}. An in-depth experimental analysis of plane-stress wave propagation -- in a specimen of similar dimensions and material properties, is provided in~\cite{pour2018} where full-field waveform data are analyzed within the frequency range $10\nxs-\nxs40\mbox{kHz}$. 

It should be mentioned that the differential indicators are a form of full-waveform inversion~\cite{pour2019}, and thus, they do not rely on a specific mode of propagation, nor they require any such knowledge on the nature of wave motion. In this study, however, the plane-stress approximation implies that the data inversion may be conducted in a reduced-order space involving the in-plane components of the measured wavefields as delineated in section~\ref{DI}.      
\end{rem}

\vspace{-2.5mm}
As illustrated in Fig.~\ref{Exp-sch}(b), the scanning grid $\bigcup_{i=1}^4\mathcal{G}_i$ is in the immediate vicinity of the external boundary of specimen. $\mathcal{G}_1$ (\emph{resp.}~$\mathcal{G}_3$) is centered in the mid- right (\emph{resp.}~left) edge of the sample with $27$ uniformly spaced measurement points over a span of $22$cm, while $\mathcal{G}_2$ (\emph{resp.}~$\mathcal{G}_4$) is at the top (\emph{resp.}~bottom) center of the plate involving a uniform grid of $45$ scan points over an interval of $38$cm. In light of Remark~\ref{pw}, this amounts to a spatial resolution of about 8mm for ultrasonic measurements at $30\mbox{kHz}$ in $\bxi_1\nxs$ and $\bxi_2$ directions. At every scan point, the data acquisition is conducted for a time period of 1ms at the sampling rate of 512kHz. To minimize the impact of (optical and mechanical) random noise in the system, the measurements are averaged over an ensemble of 60 realizations at each scan point. Furthermore, signal enhancement and speckle tracking were enabled to avoid signal dropouts due to surface roughness.       

\vspace{-2.5mm}
\begin{rem}\label{sg}
Note that the observation grid is consistent with common configurations in practice where only a subset of the domain's external boundary is accessible for (contact or non-contact) sensing. Recall that the differential indicators reconstruct the support of internal evolution from boundary (or far-field) data. Thus, full-field ultrasonic waveforms i.e., measurements on the entire surface of specimen are not captured in this study. An image processing scheme for anomaly detection by way of full-field measurements is provided in~\cite{pour2018}.  
\end{rem}
\vspace{-2.5mm}

\begin{figure}[!tp] \vspace*{-0mm}
\center\includegraphics[width=0.98\linewidth]{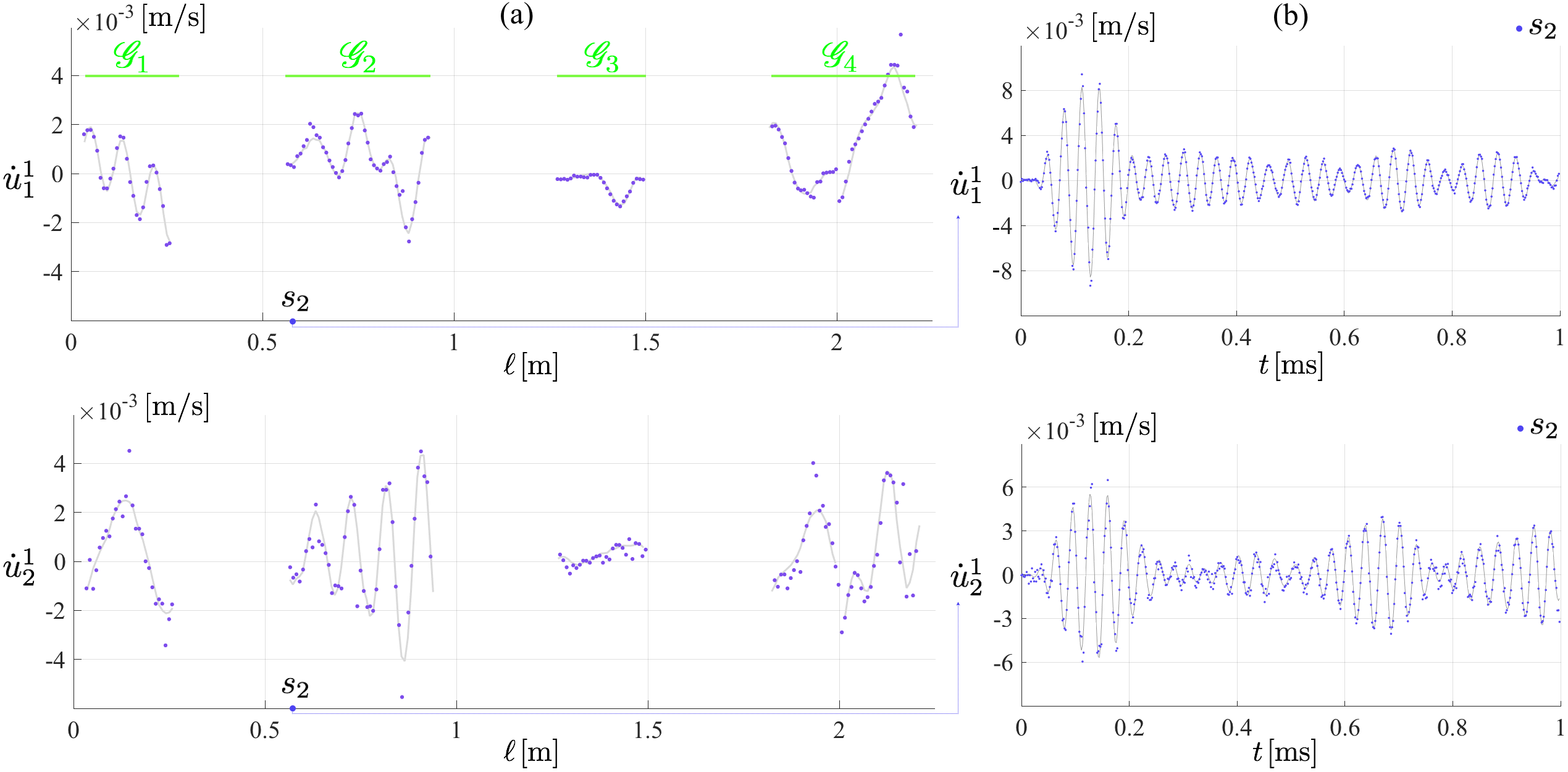} \vspace{-3.5mm}
\caption{SLDV measurements over the scanning grid $\bigcup_{i=1}^4\mathcal{G}_i$:~(a)~particle velocity distribution $\dot{u}_1^1(\ell,t=0.25\textrm{ms})$ (\emph{resp.}~$\dot{u}_2^1(\ell,t=0.25\textrm{ms})$) in $\bxi_1$ (\emph{resp.}~$\bxi_2$) direction at sensing step $t_1$, where $\ell$ represents the counterclockwise arc length along the specimen edge as in Fig.~\ref{Exp-sch}(b),~and~(b)~time history of the particle velocity response $[\dot{u}_1^1 \,\, \dot{u}_2^1](\ell=0.6\textrm{m},t)$ measured in the vicinity of transducer located at $s_2$. Dots represent ``raw" measurements and solid lines are the corresponding processed data according to section~\ref{DSP}.} \label{RS} \vspace{-5mm}
\end{figure}

To demonstrate the acquired SLDV measurements, Fig.~\ref{RS}(a) displays a snapshot in time (at $t = 0.25$ms) of the particle velocity distributions $\dot{u}_1^1$ and $\dot{u}_2^1$ over the scanning grid $\bigcup_{i=1}^4\mathcal{G}_i$ in $\bxi_1$ and $\bxi_2$ directions, respectively. These measurements are conducted at the sensing step $t_1$ when the specimen is notched with no prestressing. Note that the test data is plotted against the counterclockwise arc length $\ell$ around the specimen's external boundary whose origin is at the bottom-right corner of the plate as shown in Fig.~\ref{Exp-sch}(b). Fig.~\ref{RS}(b) plots the time history of in-plane SLDV measurements at a fixed grid point with the affiliated arc length $\ell = 0.6$m -- in the immediate vicinity of the ultrasonic source at~$s_2$ indicated in Fig.~\ref{RS}(a). It should be mentioned that in Fig.~\ref{RS}, ``raw" test data are shown with dots (corresponding to every scan grid), while the linearly interpolated solid lines show the processed data according to section~\ref{DSP}.  

\vspace{-2mm}
\begin{rem}[scattered waveforms]\label{sw}
\textcolor{black}{Recall that the differential evolution indicators rely on the spectrum of scattered field $\bv^\kappa$ which may be directly computed from the measured free field $\bu\ff$ at $t_\circ$, and total fields $\bu^\kappa$ at $t_1, \ldots, t_4$. An effort was made to generate sufficiently similar incident waveforms (up to some simple post processing measures described in section~\ref{DSP}) at each source location in all sensing steps $t_\kappa$. This is accomplished by exercising:~(i)~precise geometric alignment of the piezoelectric transducer, (ii)~application of a thin and uniform layer of cyanoacrylate glue as couplant, and (iii)~comparison of the incident waveforms captured in the vicinity of the transducer (before any reflections occur) prior to conducting the planned data acquisition.}
\end{rem}
\vspace{-2.5mm}
 
\vspace{-2.5mm}
\section{Signal processing}\label{DSP}

\noindent  This section aims to systemically extract the spectrum of scattered displacement response over the observation grid from the SLDV-measured particle velocity data. The results will be deployed in section~\ref{DI} to sequentially reconstruct the support of 3PB-induced evolution in the granite specimen. In this vein, ``raw" measurement data are processed in three steps, involving:~(1)~spatiotemporal filtering and time integration,~(2)~synchronization of incidents and extraction of scattered fields, and~(3)~spectral analysis. 

{\slshape (1)~spatiotemporal filtering and time integration.} A band-pass filter of bandwidth $20$kHz centered at $30$kHz -- consistent with the spectrum of excitation wavelet~(\ref{wavelet}), is applied to the particle-velocity records at every scan point. Note that the filtered velocity signals are temporally smooth and differentiable as shown by solid lines in Fig.~\ref{RS}(b). At every snapshot in time, however, the spatial distribution of particle velocity over the scanning grid is contaminated with data points of exceptionally low signal-to-noise ratio -- identified by sudden spikes in the observed waveforms e.g.,~see Fig.~\ref{RS}(a). To mitigate the spatial noise, first, a \emph{unified} set of observation points are specified on $\bigcup_{i=1}^4\mathcal{G}_i$ which remain invariant at all sensing steps $t_\kappa$. Then, at every time sample, four linear interpolation functions are constructed independently on $\mathcal{G}_1, \ldots, \mathcal{G}_4$ making use of (temporally filtered) velocity data points of admissible signal-to-noise ratio i.e.,~noisy points are excluded from the interpolation. In this setting, the velocity distribution at a given time may be computed over the unified observation points via the indicated interpolants. The resulting waveforms are spatially smooth as shown by solid lines in Fig.~\ref{RS}(a). It should also be mentioned that a unified observation grid across $t_\kappa$ enables arithmetic operations between data sets of distinct sensing steps, required for the computation of scattered field in step (2). Thus-obtained velocity signals are then transformed into displacement data through numerical integration. The latter process, however, introduces a low-frequency drift i.e.,~integration constant in the results, which is eliminated by a high-pass filter of cut-off frequency $500$Hz. In this way, one finds the spatiotemporally smooth ``total" displacement fields related to $\bu^\kappa$ in \eqref{uk}, which calls for further processing since the ``scattered"  fields $\bv^\kappa$ will be invoked for the imaging indicators of section~\ref{DI}.  

\begin{figure}[!tp] \vspace*{-0mm}
\center\includegraphics[width=0.98\linewidth]{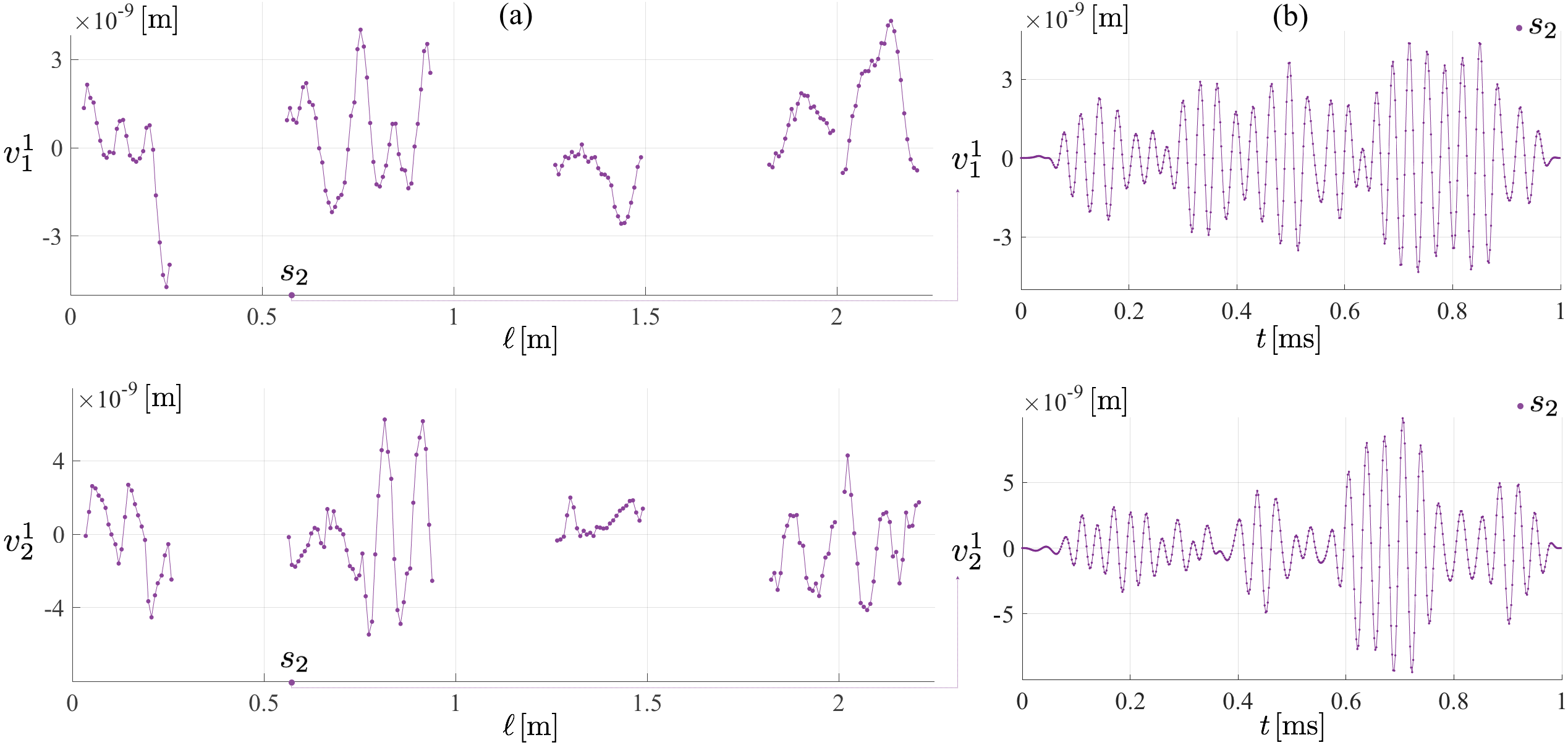} \vspace{-3.5mm}
\caption{Spatiotemporal scattered displacement field:~(a)~in-plane displacement distributions ${v}_1^1(\ell,t=0.25\textrm{ms})$ and ${v}_2^1(\ell,t=0.25\textrm{ms})$ -- in $\bxi_1\nxs$ and $\bxi_2$ directions,  at sensing step $t_1$ where $\ell$ is the arc length, and~(b)~time history of the scattered displacement response $[{v}_1^1 \,\, {v}_2^1](\ell=0.6\textrm{m},t)$ computed in the vicinity of the ultrasonic source at~$s_2$. Dots are the response affiliated with the \emph{unified} observation points, while the solid lines linearly interpolate the data points to clarify the waveforms.} \label{PS} \vspace{-4mm}
\end{figure}

{\slshape (2)~synchronization of incidents and extraction of scattered fields.}~To calculate the scattered field in light of remark~\ref{sw}, this step aims to synchronize the time, and balance the magnitude of ultrasonic incidents across $t_\kappa$. Discrepancies in transducer's physical input at various $t_\kappa$ -- although curtailed by the measures indicated in the remark, are inevitable due to~(a)~perturbation of transducer-specimen coupling in reattachments, and~(b)~recalibration of the 3D SLDV system at each $t_\kappa$. To address this problem, let us consider the displacement fields obtained in step (1) at $t_\circ$ in the vicinity of every ultrasonic source $s_1, \ldots, s_8$. The support of which is, more specifically, a subset of:~(a)~$\mathcal{G}_1$ near $s_1$, $s_8$, and $s_7$,~(b)~$\mathcal{G}_2$ in the immediate vicinity of $s_2$ and $s_3$, and~(c)~$\mathcal{G}_3$ in a neighborhood of $s_4$, $s_5$, and $s_6$. Then, the ``reference" physical incidents (transducer inputs) are identified by the first 80-100 samples of displacement time histories in the indicated neighborhoods of $s_1, \ldots, s_8$ at $t_\circ$. Note that within this timeframe i.e.,~[0 0.15]ms to [0 0.2]ms depending on the source location, there is no fingerprint on the measured waveforms due to scattering by the advancing fracture in the specimen. In this setting, the displacement fields in every ultrasonic experiment at $t_\kappa$, $\kappa \geqslant 1$, are uniformly scaled (by a constant value) and shifted in time (by a fixed amount) so that the transducer input at $t_\kappa$ matches its counterpart at $t_\circ$ i.e.,~the reference physical incident in the source location. As a result, the sequential set of ultrasonic data are consistent across all sensing steps, and one may now proceed to compute the scattered displacement fields, by subtracting the \emph{total} fields at $t_\kappa$, $\kappa \geqslant 1$ from their counterparts at $t_\circ$ as per remark~\ref{sw}. Fig.~\ref{PS} illustrates the resulting scattered field distribution in time and space at $t_1$ when the transducer is at $s_2$.   

{\slshape (3)~spectral analysis.}~This step computes the spectrum of scattered displacement signals obtained in step (2). Prior to the application of discrete Fourier transform, the problem of ``spectral leakage"~\cite{ewin1984} due to the transient nature of measured waveforms should be addressed. In this vein, the displacement data are temporally windowed~\cite{Oppe1999} using a tapered cosine i.e.~Tukey window of the form~\cite{bloo2004},   
\vspace{-1.5 mm}
\beq\label{tuw}\nonumber
\bw(\text{\bf t},\text{\sf c}) \,=\,
\!\left\{\begin{array}{l}
\vspace{0.5mm}
\!\! \dfrac{1}{2} \Big[1+\cos\nxs\big(\dfrac{2\pi}{\text{\sf c}T}(\text{\bf t}-{\text{\sf c} T}/{2})\big)\nxs\Big], \hspace{13.5mm} 0 \exs\leqslant\exs \text{\bf t} \exs<\exs  \dfrac{\text{\sf c} T}{2} \!\!\! \\*[0.5mm]
\!1, \hspace{52mm} \dfrac{\text{\sf c} T}{2} \exs\leqslant\exs \text{\bf t} \exs<\exs T-\dfrac{\text{\sf c} T}{2} \!\!\! \\*[2mm]
\!\! \dfrac{1}{2} \Big[1+\cos\nxs\big(\dfrac{2\pi}{\text{\sf c}T}(\text{\bf t}-T\!+{\text{\sf c} T}/{2})\big)\nxs\Big], \qquad T-\dfrac{\text{\sf c} T}{2} \exs\leqslant\exs \text{\bf t} \exs\leqslant\exs  T \!\!\!
\end{array}\right.
\vspace{-1.5 mm}
\eeq
where $T$ signifies the observation interval $[0 \,\,\, 1]$ms; $\text{\bf t}$ is the sampled time vector of length 512, and $0 \leqslant \text{\sf c} \leqslant 1$ is the ratio of cosine-tapered length to the entire window length. Fig.~\ref{PS}(b) shows the scattered displacements at $s_2$ after the application of Tukey window $\bw(\text{\bf t},0.2)$. Now that the support of windowed time signals is compact, one may safely proceed to compute the spectrum of scattered displacement fields via the fast Fourier transform. The resulting waveforms in the frequency domain will be used for differential imaging in section~\ref{DI}.

\vspace{-1.5mm}
\section{Data Inversion}\label{DI}

With the preceding data, one may generate the evolution indicator maps in three steps, namely by:~(i) constructing the discrete scattering operators ${\boldsymbol{\Lambda}}_\kappa$ for all $t_\kappa$, $\kappa = 1, \ldots, 4$, (ii) computing the trial signature patterns affiliated with~\eqref{PhiL}, and (iii) evaluating the differential imaging functionals~\eqref{EIFn0} through non-iterative minimization of the discretized cost functional~\eqref{GCfn}. These steps are elucidated in the following. 

\vspace{-2mm}
\subsubsection*{The discrete scattering operator}
With reference to Fig.~\ref{Exp-sch}(b), the incident surface $S^{\text{inc}}$ is sampled at $N_s = 8$ source locations $\by_j \in \lbrace s_1, s_2, \ldots, s_{8} \rbrace$, while the observation grid $S\obs = \bigcup_{\iota=1}^4\mathcal{G}_\iota$ is comprised of $N_p = 144$ measurement points $\bxi_i$. In this setting, for every $t_\kappa$, the spectrum of (in-plane) waveform data at $N_\omega = 10$ frequencies, specifically at~$\omega_\ell = 27, 28, \ldots, 36$kHz, are deployed to generate the multi-frequency scattering operator ${\boldsymbol{\Lambda}}_\kappa\nxs$ as a $2N_pN_\omega\!\times N_sN_\omega$ matrix of components     
\vspace{-1.5 mm}
\beq\lb{mat2}
{\boldsymbol{\Lambda}}_\kappa(2N_p\ell+2i+1\!:\!2N_p\ell+2i+2, \,N_s\ell+ j+1) ~=\, 
\left[\begin{array}{c}
\!\!F(v^\kappa_1)\!\!   \\*[1mm]
\!\!F(v^\kappa_2)\!\!  
\end{array}\!\right] \! (\bxi_i,\by_j;\omega_\ell),
\vspace{-1.5 mm}
\eeq
for 
\vspace{-1.5 mm}
\beq\lb{ijl}
i = 0,\ldots N_p-1, \quad j = 0,\ldots N_s-1,  \quad \ell = 0,\ldots N_\omega-1.
\vspace{-1.5 mm}
\eeq
On recalling~\eqref{So}, here, $F(v^\kappa_\iota)(\bxi_i,\by_j;\omega_\ell)$, $\iota = 1,2$, is the $\iota^{\textrm{th}}\nxs$ component of the Fourier transformed displacement at the observation point $\bxi_i$ and frequency $\omega_\ell$ when the ultrasonic source is located at $\by_j$. 

\vspace{-1.5mm}
\subsubsection*{A physics-based library of trial patterns}
Let the search volume $\mathcal{S}$ be a $29$cm $\nxs\!\times\!\nxs$ $29$cm square in the middle of specimen probed by a uniform $100 \!\times\! 100$ grid of sampling points~$\bx_{\small \circ}$ where the featured evolution indicator functionals~\eqref{EIFn0} are evaluated. In addition, the unit circle is sampled by $16$ trial normal directions $\textrm{\bf{n}}=\bR\bn_{\small \circ}$ wherein $\bn_{\small \circ}=(1,0)$. Based on this, a total of $M = 10000 \!\times\! 16$ trial dislocations $L =\bx_\circ\!+\bR{\sf L}$ are generated for the specified pairs $(\bx_{\small \circ},\textrm{\bf{n}})$. Here, ${\sf L}$ is a vertical crack of length 3mm. For each $(\bx_{\small \circ},\textrm{\bf{n}})$, the scattering signatures $\textrm{\bf{v}}^{\bx_{\small \circ\nxs},\textrm{\bf{n}}}(\bxi_i,\omega)$ are computed separately for every $\omega \in \Omega := \lbrace 27, 28, \ldots, 36 \rbrace$kHz over the observation grid $\bxi_i \in S^{\text{obs}}$ by solving
\vspace{-1.5mm}
\beq\lb{PhiL2}
\begin{aligned}
&\nabla \nxs\cdot [\bC \exs \colon \! \nabla \textrm{\bf{v}}^{\bx_{\small \circ\nxs},\textrm{\bf{n}}}](\bxi,\omega) \,+\, \rho \exs \omega^2\textrm{\bf{v}}^{\bx_{\small \circ\nxs},\textrm{\bf{n}}}(\bxi,\omega)~=~\bzero, \quad & \big(\bxi \in {\mathcal{B}}\backslash L, \omega \in \Omega \big) \\*[0.5mm]
&\bn \nxs\cdot \bC \exs \colon \!  \nabla  \textrm{\bf{v}}^{\bx_{\small \circ\nxs},\textrm{\bf{n}}}(\bxi,\omega)~=~\bzero,  \quad & \big(\bxi \in \partial{\mathcal{B}}\backslash S, \omega \in \Omega \big) \\*[0.5mm]
&\textrm{\bf{v}}^{\bx_{\small \circ\nxs},\textrm{\bf{n}}}(\bxi,\omega)~=~\bzero,  \quad & \big(\bxi \in S, \omega \in \Omega \big) \\*[0.5mm]
& \textrm{\bf{n}} \cdot \bC \exs \colon \!  \nabla  \textrm{\bf{v}}^{\bx_{\small \circ\nxs},\textrm{\bf{n}}} ~=~ |{\sf L}|^{-1} \delta (\bxi-\bx_{\small \circ}\!) \exs \textrm{\bf{n}}. \quad & \big(\bxi \in L, \omega \in \Omega \big) \\
\end{aligned}     
\vspace{-0.0mm}
\eeq

Here, $\mathcal{B}$ represents the granite specimen, and $S$ is comprised of three points where, as shown in Fig.~\ref{Exp-sch}, the two supporting pins of the load frame at the bottom and the loading pin on top meet the sample. 

These simulations are performed in three dimensions for the granite plate via an elastodynamics code rooted in the boundary element method~\cite{Bon1999, Fatemeh2015}. For data inversion, however, only the in-plane components of the computed scattered fields are used in the following form 
\vspace{-1.5 mm}
\beq\lb{Phi-inf-Dnum}
\bPhi_{\bx_{\small \circ},\textrm{\bf{n}}}(2N_p\ell+2i+1\!:\!2N_p\ell+2i+2) ~=~ 
\! \left[\begin{array}{c}  \!\! \textrm{{v}}_1^{\bx_{\small \circ\nxs},\textrm{\bf{n}}} \!\!\nxs \\*[1mm]
\!\! \textrm{{v}}_2^{\bx_{\small \circ\nxs},\textrm{\bf{n}}} \!\!\nxs
\end{array} \right]\!\nxs (\bxi_i;\omega_\ell), \qquad  i = 0,\ldots N_p-1,  \quad \ell = 0,\ldots N_\omega-1,
\vspace{-1.5 mm}
\eeq
where $\bPhi_{\bx_{\small \circ},\textrm{\bf{n}}}$ is a $2N_pN_\omega\!\times\! 1$ vector. In this setting, the scattering equation~\eqref{FF} may be discretized as
\vspace{-1.5 mm}
\beq\lb{Dff}
{\boldsymbol{\Lambda}}_\kappa \, \bg^\kappa_{\bx_{\small \circ\nxs},\textrm{\bf{n}}}~=~\bPhi_{\bx_{\small \circ},\textrm{\bf{n}}}. 
\vspace{-1.5 mm}
\eeq 

 \begin{rem}
 It is worth noting that $\bPhi_{\bx_{\small \circ},\textrm{\bf{n}}}$ is invariant with respect to the sensing steps $t_\kappa$. Hence, for computational efficiency, one may generate a $2N_pN_\omega \!\times\! M$ matrix $\bPhi$,
 \vspace{-1.5 mm}
\beq\lb{Phi-inf-Dnum2}\nonumber
\bPhi(2N_p\ell+2i+1\!:\!2N_p\ell+2i+2,m) ~=~ 
\! \left[\begin{array}{c}  \!\! \textrm{\emph{v}}_1^{(\bx_{\small \circ\nxs},\textrm{\bf{n}})_m} \!\!\! \\*[1mm]
\!\! \textrm{\emph{v}}_2^{(\bx_{\small \circ\nxs},\textrm{\bf{n}})_m} \!\!\!
\end{array} \right]\!\! (\bxi_i;\omega_\ell), \qquad  i = 0,\ldots N_p-1,  \quad \ell = 0,\ldots N_\omega-1,
\vspace{-1.5 mm}
\eeq
as the right hand side of scattering equation (\ref{Dff}) -- encompassing all choices of trial pairs $(\bx_{\small \circ},\textrm{\bf{n}})_m$, $m = 1, 2, \ldots M,$ so that one may construct the indicator maps at once for every $t_\kappa$.       
\vspace{-1.5mm}
 \end{rem}

\vspace{-1.5mm}
\subsubsection*{Differential indicators of evolution}
The scattering equation~(\ref{Dff}) may be ill-posed at all sensing steps due to~{(a)}~nonlinear nature of the inverse problem,~{(b)}~limited excitation and sensing apertures, and~{(c)}~local (e.g., interfacial) modes of wave motion -- in a neighborhood of the advancing fracture~\cite{Pyr1987} -- whose signature may not be found on~$S\obs$. Accordingly, (\ref{Dff}) will be solved via a careful regularization process by minimizing the discretized cost functional~\eqref{GCfn}. Following~\cite{pour2019}, on setting $\chi \simeq 0$, the minimizer $\bg^\kappa_{\bx_{\small \circ\nxs},\textrm{\bf{n}}}$ of~\eqref{GCfn} is computed non-iteratively by solving 
\vspace{-1 mm}
\beq \lb{min-DRJ} 
\Big( {\boldsymbol{\Lambda}}_\kappa^{\! *}{\boldsymbol{\Lambda}}_\kappa  + \gamma^{\kappa}_{\bx_{\small \circ\nxs},\textrm{\bf{n}}} \exs  ({\boldsymbol{\Lambda}}_\kappa^{\! *}{\boldsymbol{\Lambda}}_\kappa)^{\nxs\frac{1}{4}*} ({\boldsymbol{\Lambda}}_\kappa^{\! *}{\boldsymbol{\Lambda}}_\kappa)^{\nxs\frac{1}{4}} + \delta_{\kappa} \exs \gamma^{\kappa}_{\bx_{\small \circ\nxs},\textrm{\bf{n}}}  \exs \boldsymbol{I}_{N_sN_\omega\times N_sN_\omega} \Big) \exs \bg^\kappa_{\bx_{\small \circ\nxs},\textrm{\bf{n}}}  ~=~  {\boldsymbol{\Lambda}}_\kappa^{\! *} \bPhi_{\bx_{\small \circ},\textrm{\bf{n}}},
\vspace{-1 mm}
\eeq
where $(\cdot)^*$ is the Hermitian operator, $\delta_{\kappa} = 0.15\norms{\!{\boldsymbol{\Lambda}}_\kappa\!}$ indicates the estimated magnitude of noise in data, and following~\cite{Fatemeh2017},   
\vspace{-1 mm}  
\beq\lb{Alph}
\gamma^{\kappa}_{\bx_{\small \circ\nxs},\textrm{\bf{n}}} \,\, \colon \!\!\! = \,\, \frac{\eta^{\kappa}_{\bx_{\small \circ\nxs},\textrm{\bf{n}}}}{\norms{\!{\boldsymbol{\Lambda}}_\kappa\!} + \,\, \delta_\kappa}.
\vspace{-1 mm}
\eeq
Here $\eta^{\kappa}_{\bx_{\small \circ\nxs},\textrm{\bf{n}}}$ is a regularization parameter computed via the Morozov discrepancy principle~\cite{Kress1999}. As a result, $\bg^\kappa_{\bx_{\small \circ\nxs},\textrm{\bf{n}}}$ is a $N_sN_\omega\times 1$ vector (or $N_sN_\omega\times M$ matrix for all the constructed right hand sides) identifying the distribution of wavefront densities over $S^{\nxs\text{inc}\nxs}$ at sensing step $t_{\kappa}$. On repeating~\eqref{min-DRJ} for all sensing steps i.e.,~$\kappa = \lbrace 1, \ldots, 4 \rbrace$, one obtains all the necessary components to construct a the differential evolution indicator maps.   

In this vein, let us first evaluate the invariant functional
\vspace{-1.5mm}
\beq \lb{Inv2N}
\textrm{\bf I}_{\kappa}(\bg^\kappa_{\bx_{\small \circ\nxs},\textrm{\bf{n}}}, \bg^{\kappa+1}_{\bx_{\small \circ\nxs},\textrm{\bf{n}}}) = \big(\, \bg^{\kappa+1}_{\bx_{\small \circ\nxs},\textrm{\bf{n}}}-\,\bg^\kappa_{\bx_{\small \circ\nxs},\textrm{\bf{n}}},\, \boldsymbol{\Upsilon}_\kappa(\bg^{\kappa+1}_{\bx_{\small \circ\nxs},\textrm{\bf{n}}}-\,\bg^\kappa_{\bx_{\small \circ\nxs},\textrm{\bf{n}}})\big) + \exs\delta_\kappa\! \norms{\!\bg^{\kappa+1}_{\bx_{\small \circ\nxs},\textrm{\bf{n}}}-\,\bg^\kappa_{\bx_{\small \circ\nxs},\textrm{\bf{n}}}\!}^2, \quad \boldsymbol{\Upsilon}_\kappa = ({\boldsymbol{\Lambda}}_\kappa^{\! *}{\boldsymbol{\Lambda}}_\kappa)^{\nxs\frac{1}{2}}.
\vspace{-1.5mm}
\eeq
Whereby, the differential imaging functionals may be computed as follows
\vspace{-1.5mm}
\beq\lb{EIFn}
\begin{aligned}
& \textrm{\bf D}^{\kappa}_{\bx_{\small \circ\nxs},\textrm{\bf{n}}}\big(\bg^\kappa_{\bx_{\small \circ\nxs},\textrm{\bf{n}}}, \bg^{\kappa+1}_{\bx_{\small \circ\nxs},\textrm{\bf{n}}}\big) := \frac{1}{\sqrt{\textrm{\bf I}_{\kappa+1}\big(\bzero, \bg^{\kappa+1}_{\bx_{\small \circ\nxs},\textrm{\bf{n}}}\big) \Big{[}1+ \textrm{\bf I}_{\kappa+1}\big(\bzero, \bg^{\kappa+1}_{\bx_{\small \circ\nxs},\textrm{\bf{n}}}\big)\exs {\textrm{\bf I}_{\kappa}^{-1}\big(\bg^\kappa_{\bx_{\small \circ\nxs},\textrm{\bf{n}}}, \bg^{\kappa+1}_{\bx_{\small \circ\nxs},\textrm{\bf{n}}}\big)}\Big{]}}},  \\
& \tilde{\textrm{\bf D}}^{\kappa}_{\bx_{\small \circ\nxs},\textrm{\bf{n}}}\big(\bg^\kappa_{\bx_{\small \circ\nxs},\textrm{\bf{n}}}, \bg^{\kappa+1}_{\bx_{\small \circ\nxs},\textrm{\bf{n}}}\big) := \frac{1}{\sqrt{\textrm{\bf I}_{\kappa}\big(\bg^{\kappa}_{\bx_{\small \circ\nxs},\textrm{\bf{n}}}, \bzero \big) + \textrm{\bf I}_{\kappa+1}\big(\bzero, \bg^{\kappa+1}_{\bx_{\small \circ\nxs},\textrm{\bf{n}}}\big) \Big{[}1+ \textrm{\bf I}_{\kappa}\big(\bg^{\kappa}_{\bx_{\small \circ\nxs},\textrm{\bf{n}}}, \bzero\big)\exs {\textrm{\bf I}_{\kappa}^{-1}\big(\bg^\kappa_{\bx_{\small \circ\nxs},\textrm{\bf{n}}}, \bg^{\kappa+1}_{\bx_{\small \circ\nxs},\textrm{\bf{n}}}\big)}\Big{]}}}.
\end{aligned}
\vspace*{-1.5mm}
\eeq
Then, upon introducing
\beq
\big(\bg^\kappa_{\bx_{\small \circ\nxs}}, \bg^{\kappa+1}_{\bx_{\small \circ\nxs}}\big) \,\,\colon \!\!= \,\, \text{argmin}_{(\bg^\kappa_{\bx_{\small \circ\nxs},\textrm{\bf{n}}}, \bg^{\kappa+1}_{\bx_{\small \circ\nxs},\textrm{\bf{n}}})} \textrm{\bf D}^{\kappa}_{\bx_{\small \circ\nxs},\textrm{\bf{n}}}, \quad \big(\tilde{\bg}^\kappa_{\bx_{\small \circ\nxs}}, \tilde{\bg}^{\kappa+1}_{\bx_{\small \circ\nxs}}\big) \,\,\colon \!\!= \,\, \text{argmin}_{(\bg^\kappa_{\bx_{\small \circ\nxs},\textrm{\bf{n}}}, \bg^{\kappa+1}_{\bx_{\small \circ\nxs},\textrm{\bf{n}}})} \tilde{\textrm{\bf D}}^{\kappa}_{\bx_{\small \circ\nxs},\textrm{\bf{n}}},
\eeq
one obtains the indicator maps
\vspace{-1.5mm}
\beq\lb{EIFN}
\begin{aligned}
& \textrm{\bf D}_{\kappa}\big(\bg^\kappa_{\bx_{\small \circ\nxs}}, \bg^{\kappa+1}_{\bx_{\small \circ\nxs}}\big) := \frac{1}{\sqrt{\textrm{\bf I}_{\kappa+1}\big(\bzero, \bg^{\kappa+1}_{\bx_{\small \circ\nxs}}\big) \Big{[}1+ \textrm{\bf I}_{\kappa+1}\big(\bzero, \bg^{\kappa+1}_{\bx_{\small \circ\nxs}}\big)\exs {\textrm{\bf I}_{\kappa}^{-1}\big(\bg^\kappa_{\bx_{\small \circ\nxs}}, \bg^{\kappa+1}_{\bx_{\small \circ\nxs}}\big)}\Big{]}}},  \\
& \tilde{\textrm{\bf D}}_{\kappa}\big(\tilde{\bg}^\kappa_{\bx_{\small \circ\nxs}}, \tilde{\bg}^{\kappa+1}_{\bx_{\small \circ\nxs}}\big) := \frac{1}{\sqrt{\textrm{\bf I}_{\kappa}\big(\tilde{\bg}^{\kappa}_{\bx_{\small \circ\nxs}}, \bzero \big) + \textrm{\bf I}_{\kappa+1}\big(\bzero, \tilde{\bg}^{\kappa+1}_{\bx_{\small \circ\nxs}}\big) \Big{[}1+ \textrm{\bf I}_{\kappa}\big(\tilde{\bg}^{\kappa}_{\bx_{\small \circ\nxs}}, \bzero\big)\exs {\textrm{\bf I}_{\kappa}^{-1}\big(\tilde{\bg}^\kappa_{\bx_{\small \circ\nxs}}, \tilde{\bg}^{\kappa+1}_{\bx_{\small \circ\nxs}}\big)}\Big{]}}}.
\end{aligned}
\vspace*{-1.5mm}
\eeq

Here, $\textrm{\bf D}_{\kappa}$ and $\tilde{\textrm{\bf D}}_{\kappa}$ canvas the support of geometric and interfacial evolution that occur between successive sensing steps $t_\kappa$ and $t_{\kappa+1}$. More specifically, $\textrm{\bf D}_{\kappa}$ assumes its highest values at the sampling points that meet the support of newly developed or elastically evolved interfaces $\hat{\Gamma}_{\kappa+1}\! \cup \tilde{\Gamma}_{\kappa+1}$, while remaining near zero everywhere else including the loci of pre-existing scatterers within $[t_\kappa\,\, t_{\kappa+1}]$ i.e., ${\Gamma}_{\!\circ}\nxs \cup {\Gamma}_{\kappa} \backslash \tilde{\Gamma}_{\kappa+1}$. On the other hand, $\tilde{\textrm{\bf D}}_{\kappa}$ is by design sensitive to mechanical evolution achieving its most pronounced values when $\bx_{\small \circ\nxs}$ approaches $\tilde{\Gamma}_{\kappa+1}$, while assuming near zero values when $\bx_{\small \circ\nxs} \in \mathcal{S}\backslash \tilde{\Gamma}_{\kappa+1}$.   

\vspace{-1.5mm}
\section{Results and discussion}\lb{RE}

For clarity of discussion, let us recall the damage configuration at every sensing step $t_\kappa$, $\kappa = \circ, 1, \ldots, 4$. The specimen is nominally intact at $t_\circ$, i.e., $\Gamma_\circ = \emptyset$, while featuring a manufactured notch $\Gamma_1$ at $t_1$ according to Fig.~\ref{GTC}(b). $t_1$ also coincides with the onset of fracturing and the beginning of differential imaging. 
\vspace*{-1.5mm}
\begin{rem}\lb{Base}
The baseline model -- encompassing our a priori knowledge of specimen used for data inversion, consists of the geometry of intact specimen (prior to notching) and its elastodynamic properties. Thus, in what follows, $\Gamma_1$ is deemed a pre-existing scatterer at $t_1$ of unknown support. The latter assumption reflects a common situation in practice where a component (e.g., in a nuclear power plant) at the outset of ultrasonic testing feature a network of unknown scatterers due to aging. In this setting, while reconstruction of the entire component may be pursued, the primary interest is often in spatiotemporal tracking of its active process zones.     
\vspace*{-6mm}
\end{rem} 

At $t_2$, $t_3$, and $t_4$ -- when the applied load reaches, respectively, to nearly $90$\%, $75$\%, and $60$\% of its maximum value in the post-peak regime, an invisible damage zone is advancing in the specimen. For verification purposes, an attempt was made to expose the footprints of damage by spraying acetone on the back of specimen in a neighborhood of the pre-manufactured notch. While evaporating, acetone reveals the support of 3PB-induced damage as shown in Fig.~\ref{GTC}(a). Thus-captured traces at $t_\kappa$, $\kappa = 2,3,4$, are used to approximate the ``true" support of $\Gamma_\kappa$ as illustrated in Fig.~\ref{GTC}(b). These results are then compared, in Fig.~\ref{GTC}(c), with the reconstructed support of (geometric and mechanical) evolution $\hat{\Gamma}_{\kappa}\! \cup \tilde{\Gamma}_{\kappa}$ obtained via differential indicators in Fig.~\ref{TEIF} for successive timeframes $[t_{\kappa-1} \,\,\, t_{\kappa}]$.         

\begin{figure}[!h]
\center\includegraphics[width=0.59\linewidth]{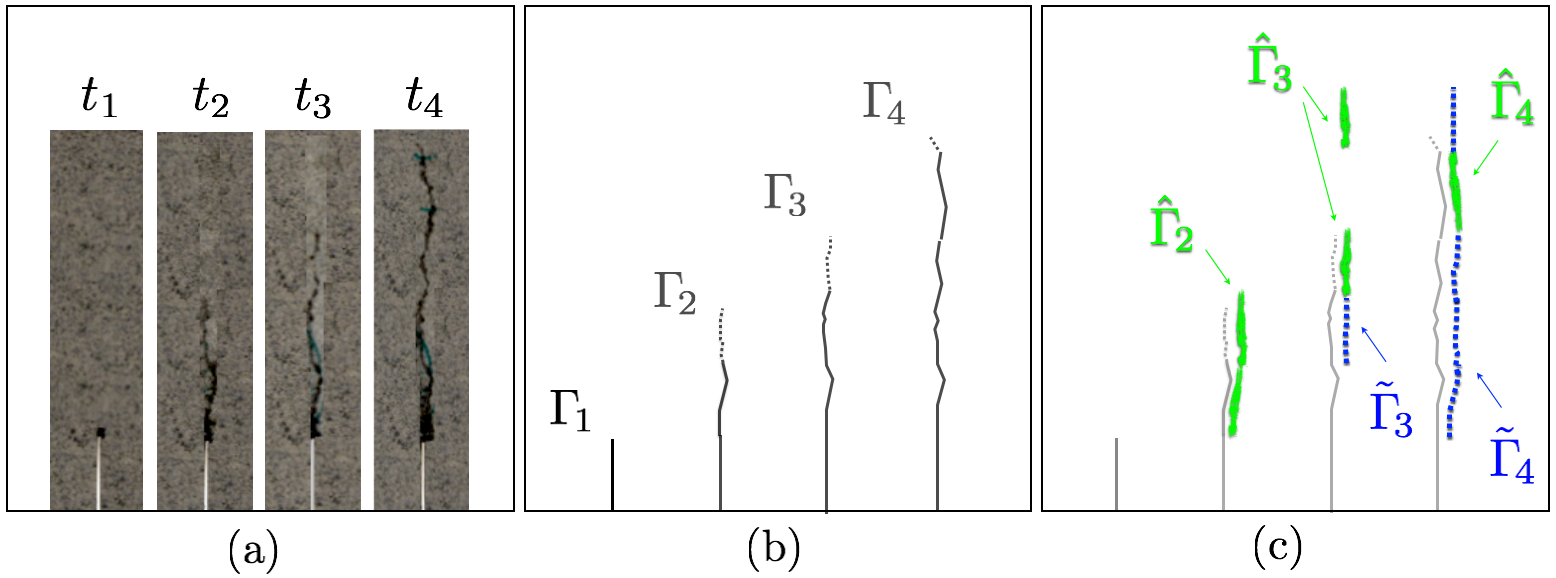} \vspace*{-3.5mm} 
\caption{The 3PB-induced fracture evolution:~(a)~damage footprints traced by acetone in a neighborhood of the pre-manufactured notch at sensing steps $t_\kappa$, $\kappa = 1,\ldots,4$,~(b)~support of $\Gamma_\kappa$ retrieved from~(a) where weak traces are identified by dashed lines, and~(c) reconstructed support of newborn fractures $\hat{\Gamma}_{\kappa+1}$ (solid lines) and mechanically evolved interfaces $\tilde{\Gamma}_{\kappa+1}$ (dashed lines) by way of the differential indicators $\textrm{\bf D}_{\kappa}\!$ and $\tilde{\textrm{\bf D}}_{\kappa}\!$ in three consecutive timeframes $[t_\kappa \,\,\, t_{\kappa+1}]$, $\kappa = 1,2,3$. Here, the recovered evolution maps are compared with the observed traces in panel (b).}\lb{GTC}
\vspace*{-2mm}
\end{figure} 

\begin{figure}[!h]
\center\includegraphics[width=0.62\linewidth]{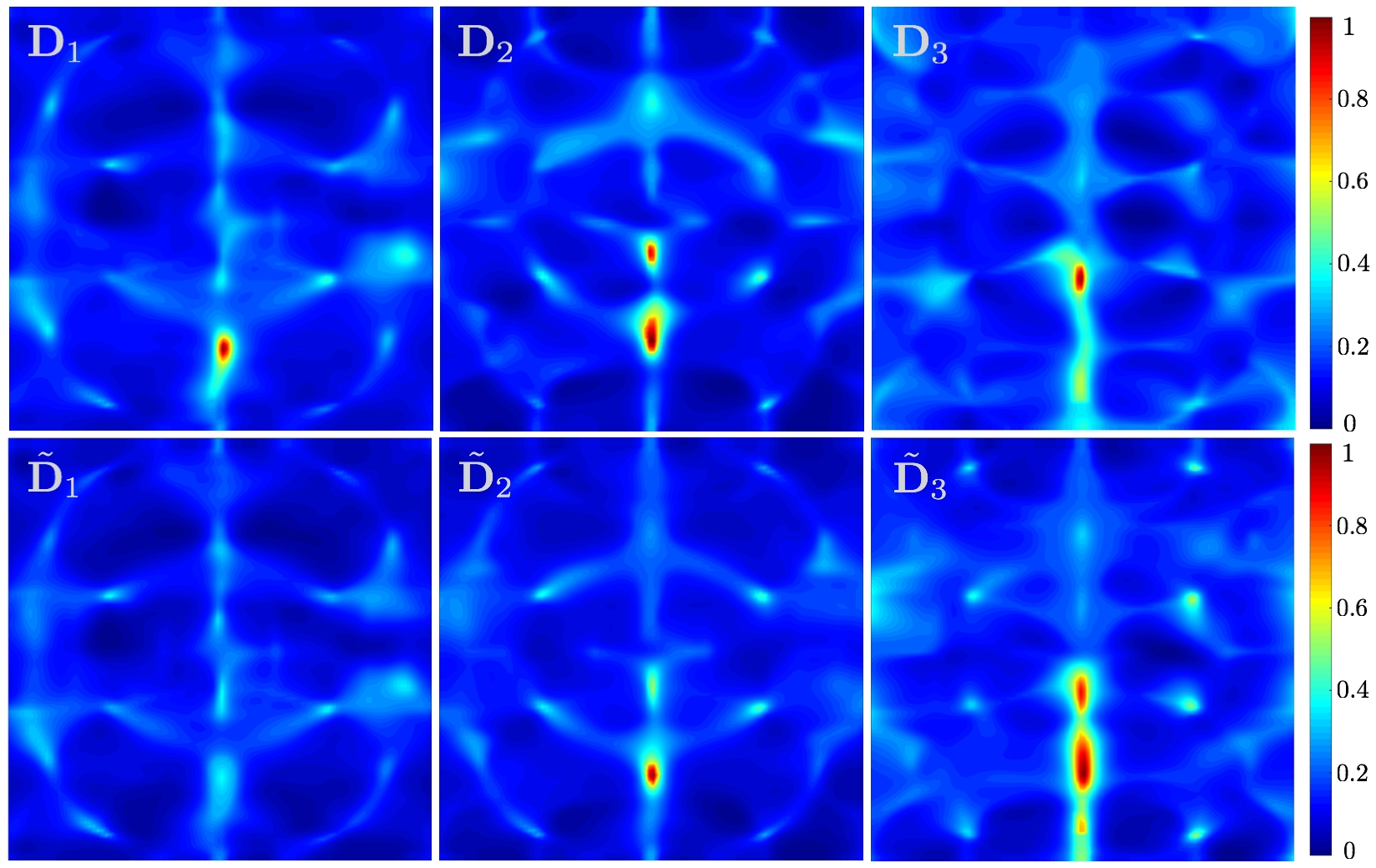} \vspace*{-2.5mm} 
\caption{Differential evolution indicator maps $\textrm{\bf D}_{\kappa}\!$ (top row) and $\tilde{\textrm{\bf D}}_{\kappa}\!$ (bottom row) computed according to~\eqref{EIFN} for $\kappa = 1,2,3$ in the sampling region -- a $29$cm $\nxs\!\times\!\nxs$ $29$cm square in the middle of specimen. $\textrm{\bf D}_{\kappa}\!$ assumes its highest values in the vicinity of newborn fractures $\hat{\Gamma}_{\kappa+1}$ and elastically evolved interfaces  $\tilde{\Gamma}_{\kappa+1}$ within the timeframe $[t_{\kappa} \,\,\, t_{\kappa+1}]$, while $\tilde{\textrm{\bf D}}_{\kappa}\!$ is primarily sensitive to mechanical i.e.,~elastic evolution and reconstruct the support of $\tilde{\Gamma}_{\kappa+1}$. Here, full ultrasonic data is deployed for the reconstruction according to Fig.~\ref{Exp-sch}(b) where $S^{\text{inc}} = \lbrace {s}_1, s_2, \ldots, s_8 \rbrace$ and $S\obs = \bigcup_{i=1}^4\mathcal{G}_i$ involving 144 measurement points.} \lb{EIF}
\vspace*{-4.5mm}
\end{figure} 

\begin{figure}[!h]
\center\includegraphics[width=0.59\linewidth]{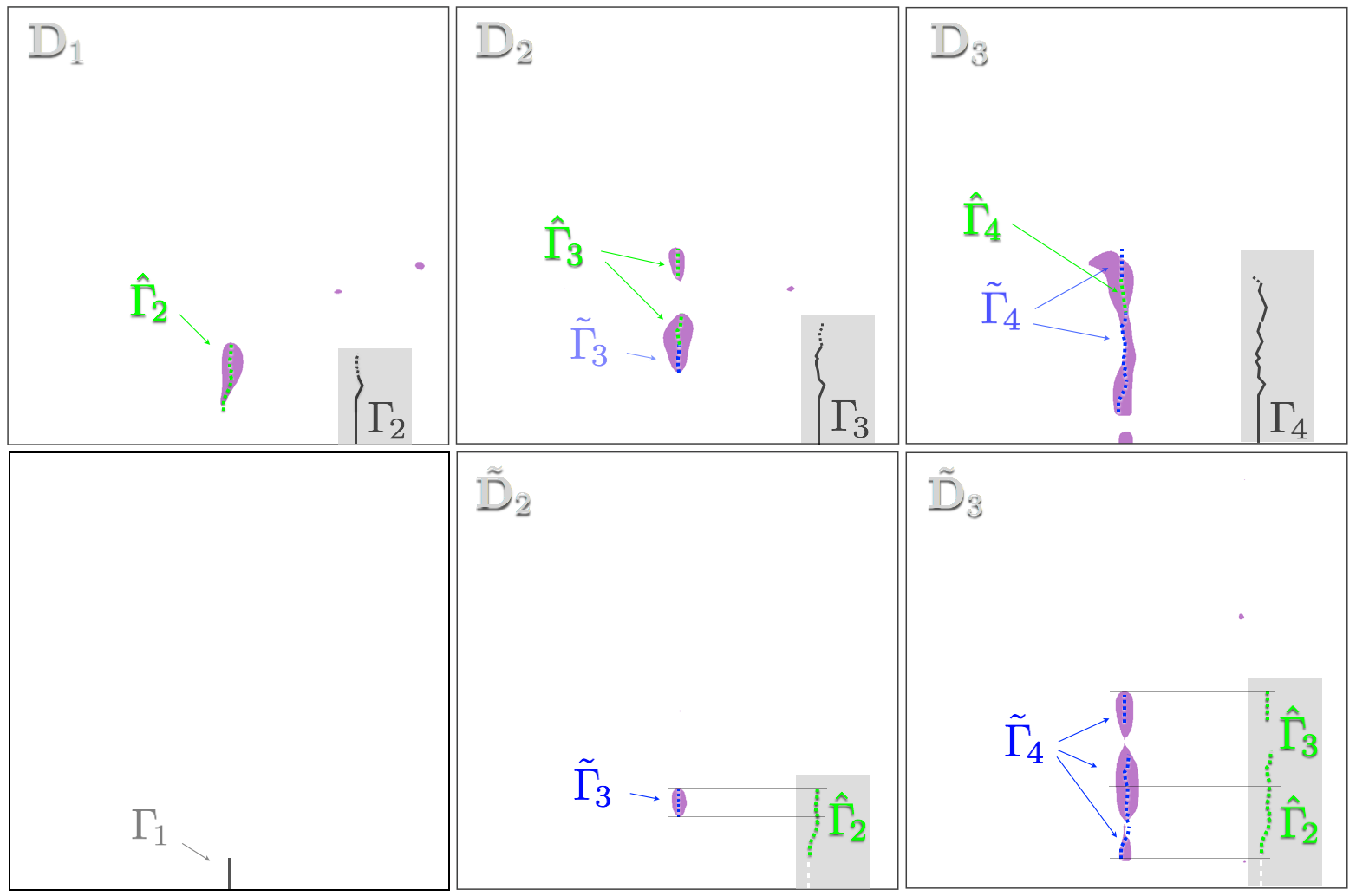} \vspace*{-2.5mm} 
\caption{Thresholded indicator maps demonstrating the loci of sampling points $\bx_{\circ\nxs}$ in Fig.~\ref{EIF} that satisfy $\textrm{\bf D}_{\kappa}(\bx_{\circ\nxs}) \geqslant \alpha \!\times\! \max(\textrm{\bf D}_{\kappa})$ (top row) and $\tilde{\textrm{\bf D}}_{\kappa}(\bx_{\circ\nxs}) \geqslant \alpha \!\times\! \max(\tilde{\textrm{\bf D}}_{\kappa})$ (bottom row) where $\alpha \in [0.55 \,\,\, 0.6]$. These plots are used to approximate the support of $\hat{\Gamma}_{\kappa+1} \cup \tilde{\Gamma}_{\kappa+1}\!$ for $\kappa = 1,2,3$. The top-row insets show the ``true" boundary of ${\Gamma}_{\kappa+1}\!$ from Fig.~\ref{Exp-sch}(b), while the insets in the bottom row display the newborn interfaces $\hat{\Gamma}_{\kappa}$ identified from $\textrm{\bf D}_{\kappa}$ maps of top row in the previous sensing sequence.} \lb{TEIF}
\vspace*{-4.5mm}
\end{figure} 

\vspace*{-1.5mm}
\subsubsection*{Full aperture reconstruction}

Consecutive pairs of scattered displacement data $\big(F(\bv^{\kappa}), F(\bv^{\kappa+1})\big)(\bxi_i,\omega_\ell)$ measured at 144 observation points $\bxi_i \in S\obs \!=\! \bigcup_{\iota=1}^4\mathcal{G}_\iota$, $i = 0, \ldots, 143$, at ten frequencies $\omega_\ell = 27, 28, \ldots, 36$kHz -- for eight source locations on $S^{\text{inc}} = \lbrace {s}_1, s_2, \ldots, s_8 \rbrace$, are deployed to compute the differential imaging functionals $\textrm{\bf D}_{\kappa}\!$ and $\tilde{\textrm{\bf D}}_{\kappa}\!$ for $\kappa = 1,2,3$. Recall that, here, the sampling region is a $29$cm $\nxs\!\times\!\nxs$ $29$cm square in the middle of specimen. The resulting evolution maps are shown in Fig.~\ref{EIF}. As mentioned earlier, $\textrm{\bf D}_{\kappa}\!$ assumes its highest values in the vicinity of newborn fractures $\hat{\Gamma}_{\kappa+1}$ and elastically evolved interfaces  $\tilde{\Gamma}_{\kappa+1}$ in the timeframe $[t_{\kappa} \,\,\, t_{\kappa+1}]$, while $\tilde{\textrm{\bf D}}_{\kappa}\!$ is primarily sensitive to mechanical evolution and reconstruct the support of $\tilde{\Gamma}_{\kappa+1}$. 
\vspace*{-1.5mm}
\begin{rem}\label{stat}
From CMOD records during the 3PB loading of specimen, it is observed that the pre-existing notch $\Gamma_1$ experiences a maximum expansion of $0.325$mm$\simeq 0.003 \lambda_s\!$ along its width. It is then plausible to assume that $\Gamma_1$, acting as a traction-free fracture, mostly remains stationary -- both geometrically and interfacially, within the course of ultrasonic experiments $[t_{1} \,\,\, t_{4}]$. This is evident in Fig.~\ref{EIF} where both $\textrm{\bf D}_{\kappa}\!$ and $\tilde{\textrm{\bf D}}_{\kappa}\!$ are insensitive to $\Gamma_1$ due to its time invariance, and thus, this scatterer is not reconstructed by the evolution indicators.  
\vspace*{-1.5mm}
\end{rem}

Due to the invariance of $\Gamma_1$ in light of Remark~\ref{stat}, note that the support of interfacial evolution $\tilde{\Gamma}_{2} = \emptyset$ within the sensing sequence $[t_{1} \,\,\, t_{2}]$. This may be observed from the $\tilde{\textrm{\bf D}}_{1}\!$ distribution when displayed on the same colormap scale as $\tilde{\textrm{\bf D}}_{2}\!$ and $\tilde{\textrm{\bf D}}_{3}\!$ as in Fig.~\ref{EIF}. It is worth mentioning that the caustics featured in the evolution maps of Fig.~\ref{EIF} are mostly governed by the illuminating wavelength, geometric symmetries of the domain, and the arrangement of sources and receivers. Their intensity is expected to decrease when the source and measurement aperture along with the number of sources and receivers increase~\cite{Fatemeh2015(2)}.

Next, the evolution indicators of Fig.~\ref{EIF} are thresholded at $55-60$\% furnishing the support of sampling points $\bx_{\circ\nxs}$ that satisfy $\textrm{\bf D}_{\kappa}(\bx_{\circ\nxs}) \geqslant \alpha \!\times\! \max(\textrm{\bf D}_{\kappa})$ and $\tilde{\textrm{\bf D}}_{\kappa}(\bx_{\circ\nxs}) \geqslant \alpha \!\times\! \max(\tilde{\textrm{\bf D}}_{\kappa})$ for $\alpha \in [0.55 \,\,\, 0.6]$. Shown in Fig.~\ref{TEIF}, these results are then used to approximate the support of $\hat{\Gamma}_{\kappa+1} \cup \tilde{\Gamma}_{\kappa+1}$ for $\kappa = 1,2,3$ as follows. Consider $\textrm{\bf D}_{1}$ in Fig.~\ref{TEIF} depicting the loci of damage $\hat{\Gamma}_{2} \cup \tilde{\Gamma}_{2}$ induced within $[t_{1} \,\,\, t_{2}]$, then on recalling $\tilde{\Gamma}_{2} = \emptyset$, the newborn fracture $\hat{\Gamma}_{2}$ is approximated by the midline through the reconstructed damage zone as shown in the figure. It is instructive to compare $\hat{\Gamma}_{2}$ with the ``true" fracture boundary $\Gamma_2$ from Fig.~\ref{GTC}(b) -- also included as an inset in Fig.~\ref{TEIF}. In this vein, observe that $\hat{\Gamma}_{2}$ has advanced slightly further in the specimen compared to $\Gamma_2$. This may be justified by noting that acetone -- used to recover $\Gamma_2$, detects only the sufficiently penetrable interfaces on the back of specimen which may not include the tight contacts in the near tip region. In the next sensing sequence $[t_{2} \,\,\, t_{3}]$, the thresholded map $\tilde{\textrm{\bf D}}_{2}$ identifies the active interface $\tilde{\Gamma}_{3}$ as a subset of $\hat{\Gamma}_{2}$ experiencing elastic evolution as the fracture propagates. Such knowledge of $\tilde{\Gamma}_{3}$ paves the way to specify the newborn fractures $\hat{\Gamma}_{3}$ from the thresholded image ${\textrm{\bf D}}_{2}$ in Fig.~\ref{TEIF}. Note that the support of evolution $\tilde{\Gamma}_{3}\!\cup \hat{\Gamma}_{3}$ in $[t_{2} \,\,\, t_{3}]$ is disjoint whose smaller segment is nearly 2cm $\simeq \lambda_s/5$ signifying the remarkable resolution of differential indicators -- similar to other imaging solutions rooted in the sampling methods~\cite{pour2019,Fatemeh2017,cako2016}. In the last sensing sequence $[t_{3} \,\,\, t_{4}]$, $\tilde{\Gamma}_{4}$ reconstructed by the thresholded $\tilde{\textrm{\bf D}}_{3}$ involves the entire 3PB-induced fracture from $t_1$ to $t_3$ i.e.,~$\hat{\Gamma}_{2}\! \cup \hat{\Gamma}_{3}$. This might be attributed to the 3PB loading configuration and the fact that at $t_4$ the fracture has almost reached the middle of specimen. More specifically, as the CMOD increases, the interfacial stiffness at the surface of $\hat{\Gamma}_{2}\! \cup \hat{\Gamma}_{3}$ decreases or may even vanish if the two faces of fracture separate, and such interfacial variations will be intrinsically more significant as the fractures grows further. On the other hand, the thresholded ${\textrm{\bf D}}_{3}$ map indicates that the two segments of damage zone coalesce at this stage via the new bridging segment $\hat{\Gamma}_{4}$. Finally, Fig.~\ref{GTC}(c) compares the retrieved support of evolution $\hat{\Gamma}_{\kappa+1} \cup \tilde{\Gamma}_{\kappa+1}$ for $\kappa = 1,2,3$ with the total fracture boundary $\Gamma_{\kappa+1}$ obtained via acetone tracing.

\vspace*{-1.5mm}
\subsubsection*{Reconstruction from reduced data}

To examine the performance of differential indicators with sparse data, the measurement points on $S\obs$ are spatially downsampled by a factor of nine so that only 16 data points shown in Fig.~\ref{REIF} are used for the reconstruction -- instead of 144 points used to obtain Figs.~\ref{EIF} and~\ref{TEIF}. The results are shown in~Fig.~\ref{REIF} where both indicator maps $\textrm{\bf D}_{\kappa}\!$ and $\tilde{\textrm{\bf D}}_{\kappa}$, $\kappa = 1,2,3$, appear to be successful in imaging the evolving damage zone. Compared to Fig.~\ref{EIF}, however, the caustics are more intense which is rather expected with reference to the in-depth analysis conducted in~\cite{Fatemeh2015(2)}.   

\begin{figure}[!h]
\center\includegraphics[width=0.62\linewidth]{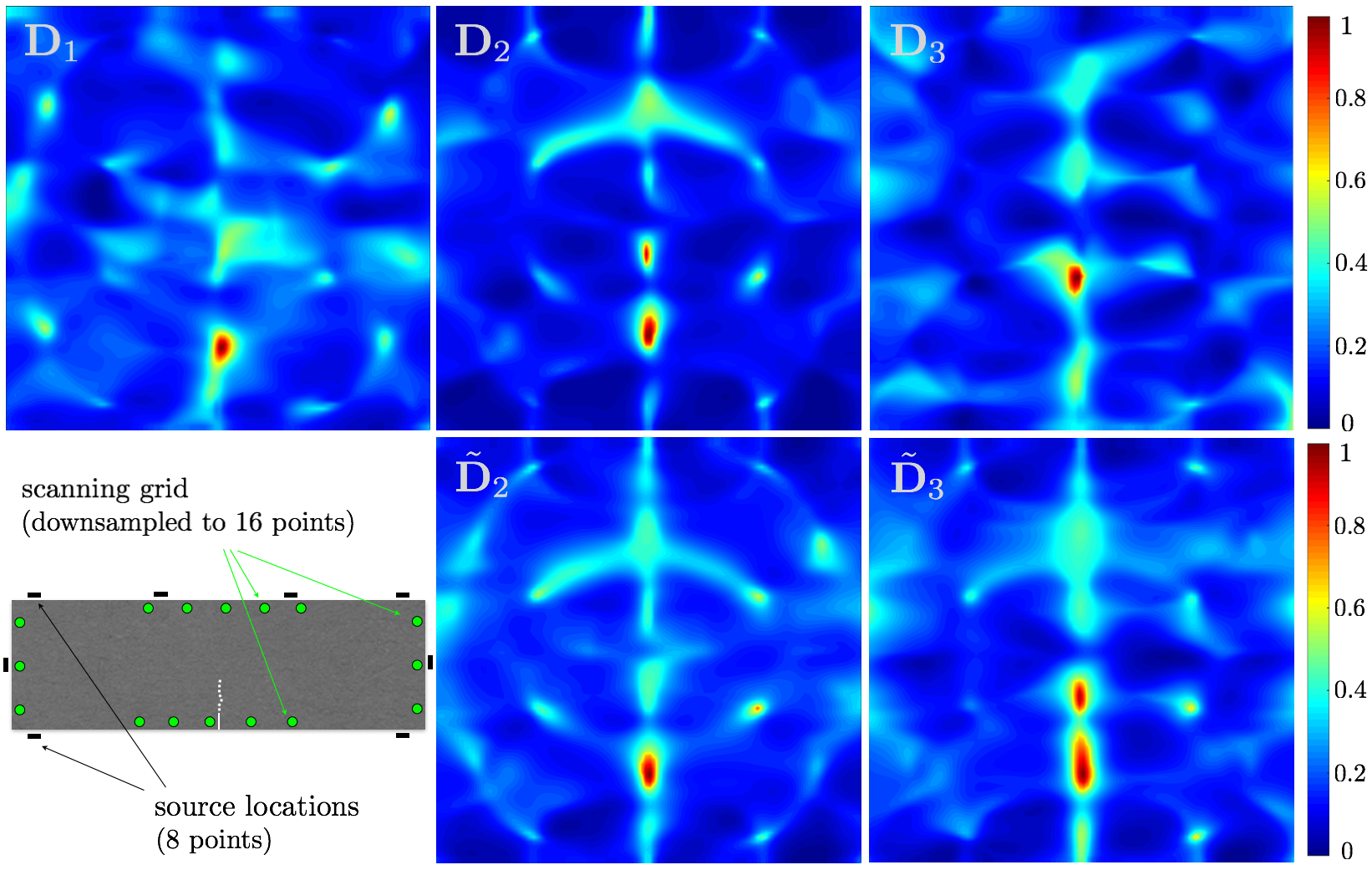} \vspace*{-2.5mm} 
\caption{Evolution indicator maps $\textrm{\bf D}_{\kappa}\!$ (top row) and $\tilde{\textrm{\bf D}}_{\kappa}\!$ (bottom row), $\kappa = 1,2,3$, constructed from reduced data where $S^{\text{inc}} = \lbrace {s}_1, s_2, \ldots, s_8 \rbrace$ and $S\obs = \bigcup_{i=1}^4\mathcal{G}_i$ involving 16 measurement points shown in the bottom left panel i.e., spatial resolution of measurements is reduced by a factor of nine.} \lb{REIF}
\vspace*{-2.5mm}
\end{figure}

\vspace*{-1.5mm}
\subsubsection*{Partial source and ``viewing" aperture}

It is common in practice that a specimen is inaccessible from one side or, to the contrary, is only accessible from one side for ultrasonic testing. Imaging in such configurations are investigated in Figs.~\ref{PEIF} and~\ref{OEIF}. More specifically, Fig.~\ref{PEIF} illustrates the differential evolution maps $\textrm{\bf D}_{\kappa}\!$ and $\tilde{\textrm{\bf D}}_{\kappa}$ for $\kappa = 1,2,3$ when the specimen is inaccessible from below for both excitation and measurement. In this setting, the reconstruction is performed using data from six source locations -- i.e., $S^{\text{inc}} = \lbrace {s}_1, s_2, s_3, s_4, s_5, s_8 \rbrace$, and measurements on three sides of the boundary $S\obs = \bigcup_{i=1}^3\mathcal{G}_i$ involving 99 points as shown in the figure.

Also, Fig.~\ref{OEIF} shows the evolution indicator maps $\textrm{\bf D}_{\kappa}\!$ and $\tilde{\textrm{\bf D}}_{\kappa}\!$, $\kappa = 1,2,3$, when the specimen is merely accessible from the top for ultrasonic illumination and sensing. In this case, indicator functionals are computed using limited data involving four ultrasonic sources on top $S^{\text{inc}} = \lbrace {s}_1, s_2, s_3, s_4 \rbrace$, and 45 measurement points on $S\obs = \mathcal{G}_2$ as shown in the figure.

\begin{figure}[!h]
\center\includegraphics[width=0.62\linewidth]{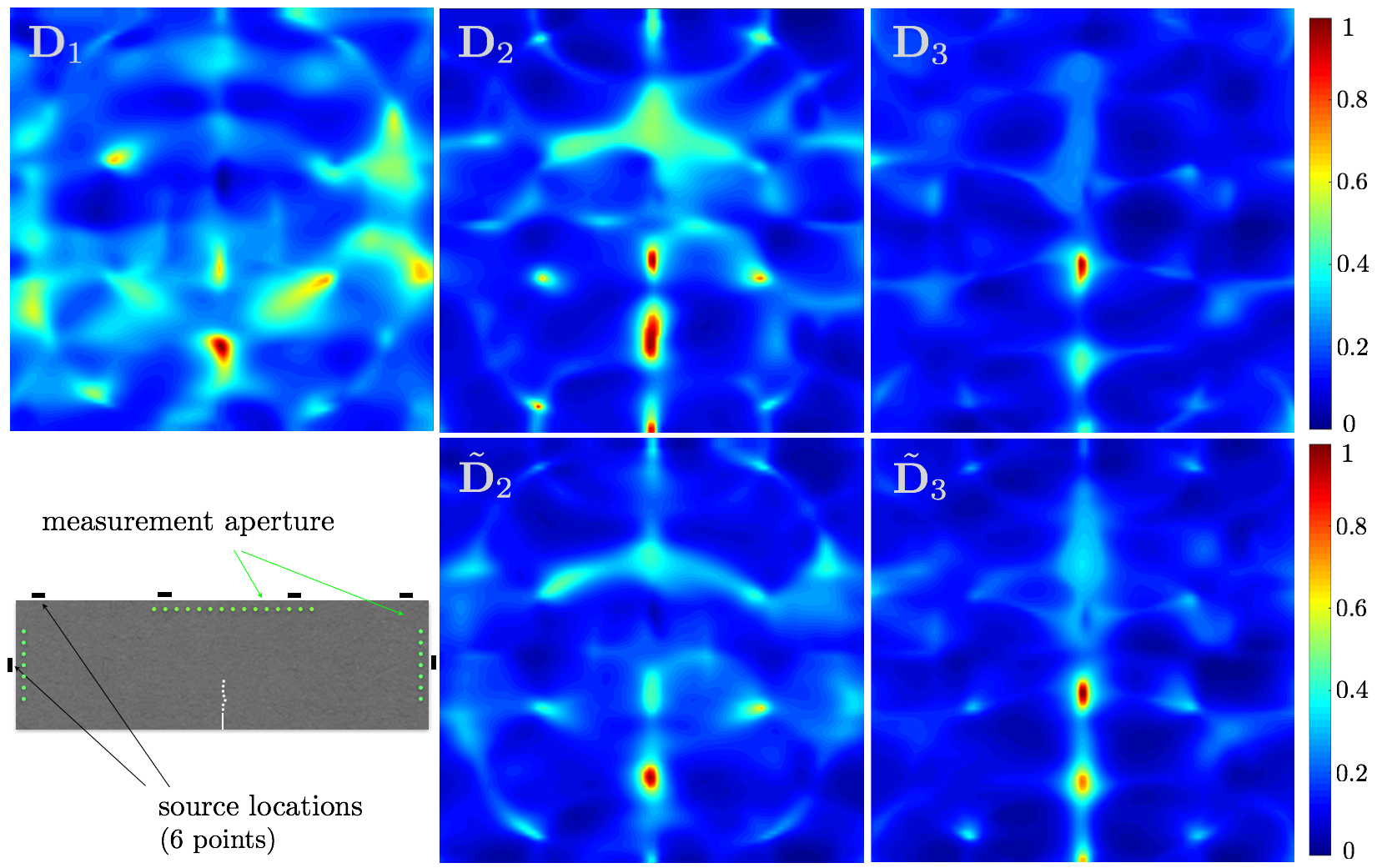} \vspace*{-2.5mm} 
\caption{Partial-aperture tomography:~differential evolution maps $\textrm{\bf D}_{\kappa}\!$ (top row) and $\tilde{\textrm{\bf D}}_{\kappa}\!$ (bottom row) constructed for $\kappa = 1,2,3$ when $S^{\text{inc}} = \lbrace {s}_1, s_2, s_3, s_4, s_5, s_8 \rbrace$ and $S\obs = \bigcup_{i=1}^3\mathcal{G}_i$ involving 99 measurement points as shown in the bottom left panel, i.e., data related to ultrasonic sources and measurement points on the bottom of specimen is ignored in the reconstruction.} \lb{PEIF}
\vspace*{-2.5mm}
\end{figure}

\begin{figure}[!h]
\center\includegraphics[width=0.62\linewidth]{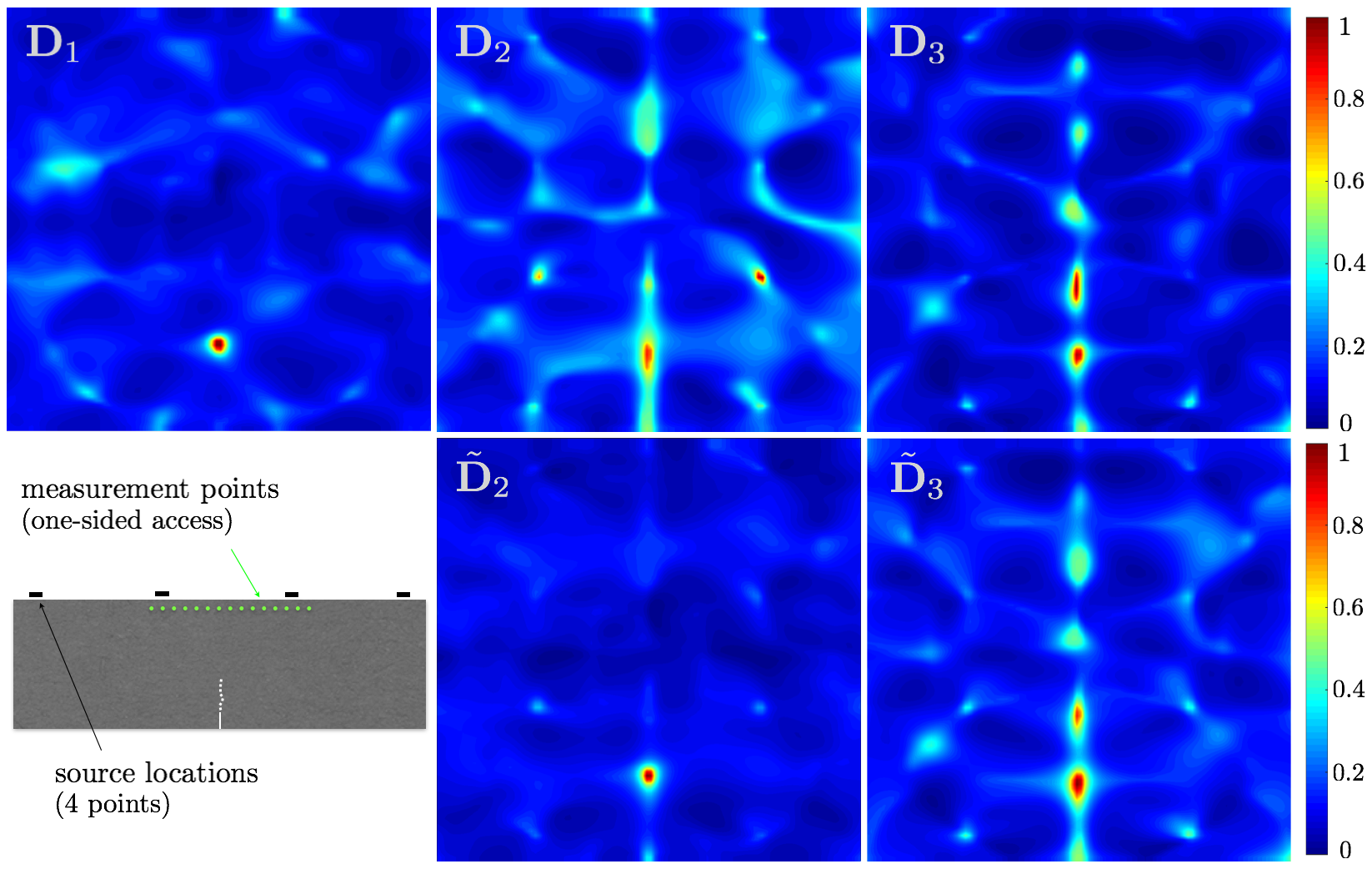} \vspace*{-2.5mm} 
\caption{One-sided reconstruction:~evolution indicator maps $\textrm{\bf D}_{\kappa}\!$ (top row) and $\tilde{\textrm{\bf D}}_{\kappa}\!$ (bottom row), $\kappa = 1,2,3$, computed using limited data involving four ultrasonic sources on top $S^{\text{inc}} = \lbrace {s}_1, s_2, s_3, s_4 \rbrace$, and 45 measurement points on $S\obs = \mathcal{G}_2$ as shown in the bottom left panel.} \lb{OEIF}
\vspace*{-2.5mm}
\end{figure}

\section{Conclusions}
 
\noindent An experimental and data analysis framework is developed for in-situ waveform tomography of damage evolution in elastic backgrounds. To this end, we take advantage of the recently established differential evolution indicators for non-iterative, full-waveform reconstruction of a propagating damage zone in a granite specimen using boundary observations of scattered ultrasonic waveforms. In this vein, transient waves ranging from 20 to 40kHz are periodically induced in the sample at four sensing steps $t_\kappa$, $\kappa = 1, \ldots, 4$, while a mode-I fracture is driven in the specimen. Thus generated velocity responses are then monitored by a 3D scanning laser Doppler vibrometer over the domain's external boundary, which upon suitable signal processing furnish the spectra of scattered displacement fields over the designated scanning grid. Such sensory data are then deployed to compute the differential indicator maps reconstructing the spatiotemporal evolution of damage zone, in terms of geometry and interfacial condition, in three consecutive timeframes $[t_\kappa \,\,\, t_{\kappa+1}]$, $\kappa = 1, 2, 3$. The recovered support of geometric evolution in each sensing sequence is verified against in-situ observations, while the interfacial evolution results are analyzed for self-consistency. The differential imaging indicators are shown to be sensitive to mechanical processes with characteristic length scale of $O(\lambda_s/5)$ promising a high-resolution reconstruction of active zones. This opens the door toward: (a) in-depth analysis of multi-scale fracture networks, including their evolution and coalescence, under various loading scenarios, and (b) better understanding of the nature of interfacial evolution and its (precursory) relation with damage propagation. It is further demonstrated that the data inversion results remain robust with reduced i.e., spatially downsampled data, as well as partial-aperture data e.g., when access to specimen for excitation and sensing is limited. Another unique opportunity provided by the present framework is that of exposing the support of evolution in a background with unknown pre-existing scatterers such as the pre-manufactured notch in this study. As a perspective, it would be interesting to implement this approach in a highly scattering specimen.

\section*{Acknowledgments} 

\noindent The experimental campaign was conducted in the Department of Civil, Environmental \& Geo- Engineering at the University of Minnesota. The author kindly acknowledges the comprehensive support provided by Professors Bojan Guzina and Joe Labuz in the course of experiments. Special thanks are due to Mateus Rodrigues Batata and Ali Tarokh for their assistance with the experiments. This study was funded by the University of Colorado Boulder through FP's startup. This work utilized resources from the University of Colorado Boulder Research Computing Group, which is supported by the National Science Foundation (awards ACI-1532235 and ACI-1532236), the University of Colorado Boulder, and Colorado State University.   

\bibliography{inverse,crackbib}

\begin{thebibliography}{10}
\expandafter\ifx\csname url\endcsname\relax
  \def\url#1{\texttt{#1}}\fi
\expandafter\ifx\csname urlprefix\endcsname\relax\def\urlprefix{URL }\fi
\expandafter\ifx\csname href\endcsname\relax
  \def\href#1#2{#2} \def\path#1{#1}\fi

\bibitem{pour2019}
F.~Pourahmadian, H.~Haddar, Differential tomography of micromechanical
  evolution in elastic materials of unknown micro/macrostructure, SIAM Journal
  on Imaging Sciences 13~(3) (2020) 1302--1330.

\bibitem{rose2014}
J.~L. Rose, Ultrasonic guided waves in solid media, Cambridge university press,
  2014.

\bibitem{matl2015}
K.~H. Matlack, J.-Y. Kim, L.~Jacobs, J.~Qu, Review of second harmonic
  generation measurement techniques for material state determination in metals,
  Journal of Nondestructive Evaluation 34~(1) (2015) 273.

\bibitem{amin2017}
M.~G. Amin, Through-the-wall radar imaging, CRC press, 2017.

\bibitem{cobl2015}
J.~Coble, P.~Ramuhalli, L.~J. Bond, J.~Hines, B.~Ipadhyaya, A review of
  prognostics and health management applications in nuclear power plants,
  International Journal of prognostics and health management 6 (2015) 016.

\bibitem{hung2013}
Y.~Y. Hung, L.~X. Yang, Y.~H. Huang, Non-destructive evaluation of composites:
  digital shearography, in: Non-Destructive Evaluation of Polymer Matrix
  Composites, Elsevier, 2013, pp. 84--115.

\bibitem{hank2016}
R.~Hanke, T.~Fuchs, M.~Salamon, S.~Zabler, X-ray microtomography for materials
  characterization, in: Materials Characterization Using Nondestructive
  Evaluation (NDE) Methods, Elsevier, 2016, pp. 45--79.

\bibitem{fuen2019}
R.~Fuentes, C.~Mineo, S.~G. Pierce, K.~Worden, E.~J. Cross, A probabilistic
  compressive sensing framework with applications to ultrasound signal
  processing, Mechanical Systems and Signal Processing 117 (2019) 383--402.

\bibitem{klos2020}
G.~K{\l}osowski, T.~Rymarczyk, K.~Kania, A.~{\'S}wi{\'c}, T.~Cieplak,
  Maintenance of industrial reactors supported by deep learning driven
  ultrasound tomography, Eksploatacja i Niezawodno{\'s}{\'c} 22~(1) (2020).

\bibitem{Fatemeh2017}
F.~Pourahmadian, B.~B. Guzina, H.~Haddar, Generalized linear sampling method
  for elastic-wave sensing of heterogeneous fractures, Inverse Problems 33~(5)
  (2017) 055007.

\bibitem{cant2019}
S.~Cantero-Chinchilla, J.~Chiach{\'\i}o, M.~Chiach{\'\i}o, D.~Chronopoulos,
  A.~Jones, A robust bayesian methodology for damage localization in plate-like
  structures using ultrasonic guided-waves, Mechanical Systems and Signal
  Processing 122 (2019) 192--205.

\bibitem{ebra2019}
A.~Ebrahimkhanlou, B.~Dubuc, S.~Salamone, A generalizable deep learning
  framework for localizing and characterizing acoustic emission sources in
  riveted metallic panels, Mechanical Systems and Signal Processing 130 (2019)
  248--272.

\bibitem{jia2016}
F.~Jia, Y.~Lei, J.~Lin, X.~Zhou, N.~Lu, Deep neural networks: A promising tool
  for fault characteristic mining and intelligent diagnosis of rotating
  machinery with massive data, Mechanical Systems and Signal Processing 72
  (2016) 303--315.

\bibitem{barn2013}
D.~Barnard, L.~J. Bond, J.~Bowler, N.~Bowler, L.~Brasche, C.~Chiou,
  A.~Frishman, J.~Gray, T.~Gray, S.~D. Holland, Quantitative inspection
  technologies for aging military aircraft, Tech. rep., Iowa State University
  Ames Center for Nondestructive Evaluation (2013).

\bibitem{ever2016}
S.~K. Everton, M.~Hirsch, P.~Stravroulakis, R.~K. Leach, A.~T. Clare, Review of
  in-situ process monitoring and in-situ metrology for metal additive
  manufacturing, Materials \& Design 95 (2016) 431--445.

\bibitem{Verdon2013b}
A.~F. Baird, J.-M. Kendall, J.~P. Verdon, A.~Wuestefeld, T.~E. Noble, Y.~Li,
  M.~Dutko, Q.~J. Fisher, Monitoring increases in fracture connectivity during
  hydraulic stimulations from temporal variations in shear wave splitting
  polarization, Geophys. J. Int. (2013) ggt274.

\bibitem{Verdon2013}
J.~P. Verdon, A.~Wustefeld, Measurement of the normal/tangential fracture
  compliance ratio $(z_{N}/z_{T})$ during hydraulic fracture stimulation using
  s-wave splitting data, Geophysical Prospecting 61 (2013) 461--475.

\bibitem{Taron2014}
J.~Taron, D.~Elsworth, Coupled mechanical and chemical processes in engineered
  geothermal reservoirs with dynamic permeability, Int. J. Rock Mech. Min. Sci.
  47 (2010) 1339--1348.

\bibitem{cako2016}
F.~Cakoni, D.~Colton, H.~Haddar, Inverse Scattering Theory and Transmission
  Eigenvalues, SIAM, 2016.

\bibitem{bonn2019}
M.~Bonnet, F.~Cakoni, Analysis of topological derivative as a tool for
  qualitative identification, Inverse Problems 35~(10) (2019) 104007.

\bibitem{audi2015}
L.~Audibert, A.~Girard, H.~Haddar, Identifying defects in an unknown background
  using differential measurements, Inverse Probl. Imaging 9~(3) (2015)
  625--643.

\bibitem{de2018}
I.~De~Teresa, F.~Pourahmadian, Real-time imaging of interfacial damage in
  heterogeneous composites, SIAM Journal on Applied Mathematics 78~(5) (2018)
  2763--2790.

\bibitem{Fatemeh2017(2)}
F.~Pourahmadian, B.~B. Guzina, H.~Haddar, A synoptic approach to the seismic
  sensing of heterogeneous fractures: From geometric reconstruction to
  interfacial characterization, Computer Methods in Applied Mechanics and
  Engineering 324 (2017) 395 -- 412.

\bibitem{baro2016}
V.~Baronian, L.~Bourgeois, A.~Recoquillay, Imaging an acoustic waveguide from
  surface data in the time domain, Wave Motion 66 (2016) 68--87.

\bibitem{baro2018}
V.~Baronian, L.~Bourgeois, B.~Chapuis, A.~Recoquillay, Linear sampling method
  applied to non destructive testing of an elastic waveguide: theory, numerics
  and experiments, Inverse Problems 34~(7) (2018) 075006.

\bibitem{Kirsch2008}
A.~Kirsch, N.~Grinberg, The Factorization Method for Inverse Problems, Oxford
  University Press, Oxford, 2008.

\bibitem{Polytec}
{Polytec, Inc.}, Basic {P}rinciples of {V}ibrometry,
  \url{https://www.polytec.com/us/vibrometry/technology/} (2020, accessed
  09/10/20).

\bibitem{Mal1969}
L.~E. Malvern, Introduction to the Mechanics of a Continuous Medium,
  Prentice-Hall, Englewood Cliffs, 1969.

\bibitem{Lamb1917}
H.~Lamb, On waves in an elastic plate, Proc. R. Soc. A 93 (1917) 114 -- 128.

\bibitem{pour2018}
F.~Pourahmadian, B.~B. Guzina, On the elastic anatomy of heterogeneous
  fractures in rock, Int J Rock Mech Min 106 (2018) 259--268.

\bibitem{ewin1984}
D.~J. Ewins, Modal testing:~theory and practice, Research studies press,
  Letchworth, 1984.

\bibitem{Oppe1999}
A.~V. Oppenheim, R.~W. Schafer, J.~R. Buck, Discrete-time signal processing,
  Prentice Hall, Upper Saddle River, NJ, 1999.

\bibitem{bloo2004}
P.~Bloomfield, Fourier analysis of time series:~an introduction, John Wiley \&
  Sons, 2004.

\bibitem{Bon1999}
M.~Bonnet, Boundary Integral Equation Methods for Solids and Fluids, Wiley,
  1999.

\bibitem{Fatemeh2015}
F.~Pourahmadian, B.~B. Guzina, On the elastic-wave imaging and characterization
  of fractures with specific stiffness, Int. J Solids Struct. 71 (2015)
  126--140.

\bibitem{Pyr1987}
L.~J. Pyrak-Nolte, N.~G.~W. Cook, Elastic interface waves along a fracture,
  Geophys. Res. Let. 14 (1987) 1107--1110.

\bibitem{Kress1999}
R.~Kress, Linear integral equation, Springer, Berlin, 1999.

\bibitem{Fatemeh2015(2)}
B.~B. Guzina, F.~Pourahmadian, Why the high-frequency inverse scattering by
  topological sensitivity may work, Proc. R. Soc. A 471 (2015) 20150187.

\end{thebibliography}

\end{document}